\documentclass[aps, prd, superscriptaddress, preprintnumbers, floatfix]{revtex4}

\usepackage{graphicx,amsfonts,amsmath,amssymb,amstext}
\usepackage{float,wrapfig}
\usepackage{subfigure, psfrag}
\usepackage{dsfont}
\usepackage{color}
\usepackage{hyperref}
\hypersetup{
	colorlinks = true,
	linkcolor = {blue},
	citecolor = {blue},
    urlcolor = {blue}
}
\usepackage{bm}

\newcommand{\be}{\begin{equation}}
\newcommand{\ee}{\end{equation}}
\newcommand{\ba}{\begin{eqnarray}}
\newcommand{\ea}{\end{eqnarray}}
\newcommand{\sign}[1]{\,\mbox{sgn}\left({#1}\right)}

\definecolor{purple}{rgb}{0.8,0,0.6}
\definecolor{darkgreen}{rgb}{0.00,0.6,0.00}

\extrafloats{100}

\begin{document}

\title{Surface plasmon polaritons in strained Weyl semimetals}
\date{August 25, 2020}

\author{O.~V.~Bugaiko}
%\email{}
\affiliation{Faculty of Physics, Kyiv National Taras Shevchenko University, 64/13 Volodymyrska st., 01601 Kyiv, Ukraine}

\author{E.~V.~Gorbar}
%\email{gorbar@bitp.kiev.ua}
\affiliation{Faculty of Physics, Kyiv National Taras Shevchenko University, 64/13 Volodymyrska st., 01601 Kyiv, Ukraine}
\affiliation{Bogolyubov Institute for Theoretical Physics, Kyiv, 03680, Ukraine}

\author{P.~O.~Sukhachov}
\email{pavlo.sukhachov@su.se}
\affiliation{Nordita, KTH Royal Institute of Technology and Stockholm University, Roslagstullsbacken 23, SE-106 91 Stockholm, Sweden}

\begin{abstract}
Surface plasmon polaritons in a strained slab of a Weyl semimetal with broken time-reversal symmetry are investigated. It is found that the strain-induced axial gauge field reduces frequencies of these collective modes for intermediate values of the wave vector. Depending on the relative orientation of the separation of Weyl nodes in momentum space, the surface normal, and the direction of propagation, the dispersion relation of surface plasmon polaritons could be nonreciprocal even in a thin slab. In addition, strain-induced axial gauge fields can significantly affect the localization properties of the collective modes. These effects allow for an {\it in situ} control of the propagation of surface plasmon polaritons in Weyl semimetals and might be useful for creating nonreciprocal devices.
\end{abstract}

\maketitle

\section{Introduction}
\label{sec:introduction}

Collective excitations are simple and informative probes of various physical properties in solids. Among them, excitations related to the interaction of light and matter are, perhaps, among the most numerous. In particular, polaritons are quasiparticles related to the coupling of electromagnetic waves with any resonance in material. The paradigmatic example of polaritons is realized by coupled states of electromagnetic waves with phonons in ionic crystals~\cite{Tolpygo,Huang} whose charged particles are not mobile. The latter property makes these materials insulating and allows for an unobstructed propagation of electromagnetic collective modes. The situation is different in metals, which are characterized by a large number of conducting electrons, where electromagnetic waves can propagate only with frequencies higher than the plasma one. Still, the surface plasmons~\cite{Ritchie:1957} can propagate with frequencies below the plasma edge. A strong interaction of light with surface plasmons produces surface plasmon polaritons (SSPs), which are, therefore, a particular case of polaritons confined to a metal-dielectric or metal-air interface.

Surface plasmon polaritons are particularly important for practical applications~\cite{Maier:book}. Indeed, they can be guided along surfaces and have significantly smaller wavelength than that of the incident photons enabling subwavelength optics and lithography beyond the diffraction limit. Further, SPPs are very sensitive to external fields, non-linear effects, and material parameters. This can be used to create nanoscale devices connected with optical switching and biosensing. Furthermore, the strong sensitivity allows one to investigate various properties of novel materials.

Recently, materials characterized by nontrivial topological properties have attracted a significant attention. A paradigmatic example of topological matter with gapless energy spectrum is given by Weyl semimetals~\cite{Yan-Felser:2017-Rev,Hasan-Huang:rev-2017,Armitage-Vishwanath:2017-Rev}. Their low-energy excitations are described by the relativistic-like Weyl equation in the vicinity of the band-touching points called Weyl nodes. Each of these nodes is a monopole of the Berry curvature, whose flux defines a topological charge of the nodes. As proved by Nielsen and Ninomiya~\cite{Nielsen-Ninomiya-1,Nielsen-Ninomiya-2}, the Weyl nodes in lattice systems always come in pairs of opposite chirality, or, equivalently, topological charges. In each pair, the Weyl nodes are separated by $2\mathbf{b}$ in momentum [this breaks the time-reversal (TR) symmetry] and/or $2b_0$ in energy (this breaks the parity-inversion symmetry). The former parameter is known as the chiral shift~\cite{Gorbar:2009bm}. It results in the anomalous Hall effect (AHE)~\cite{Yang-Lu:2011,Burkov-Balents:2011,Burkov-Hook:2011,Grushin:2012,Zyuzin:2012,Goswami:2013,Burkov:2014} in Weyl semimetals, which plays an important role in transport and optical properties of Weyl semimetals. Moreover, the AHE strongly affects collective excitations, including the SPPs.

Surface plasmon polaritons in Weyl semimetals were studied in Refs.~\cite{Zyuzin-Zyuzin:2014,Hofmann-DasSarma:2016,Kotov-Lozovik:2016,Kotov-Lozovik:2018,Tamaya-Kawabata:2019,Chen-Belyanin:2019a,Chen-Belyanin:2019b,Abdol-Abdollahipour:2019,Jalali-Mola-Jafari:2019,Abdol-Vala:2020}. The principal finding is that the SPP dispersion in Weyl semimetals with broken TR symmetry is similar to magnetoplasmons in ordinary metals~\cite{Chiu-Quinn:1972,Wallis-Hartstein:1974,Kushwaha-Halevi:1987,Boardman:book} with strong gyrotropic and nonreciprocity effects. It is important to emphasize that a giant nonreciprocity can be attained in the absence of magnetic fields, which is very advantageous for technological applications. In thin films of Weyl semimetals, a hybridization between plasmons localized at the opposite surfaces of the semimetal results in mixed plasmon modes with different localization lengths~\cite{Tamaya-Kawabata:2019}.

The nontrivial bulk topology of Weyl semimetals is also reflected in unusual surface states known as the Fermi arcs~\cite{Savrasov:2011}. Unlike surface states in ordinary materials, the Fermi arcs form open segments in momentum space that connect Weyl nodes of opposite chirality~\cite{Savrasov:2011,Haldane:2014}. The interplay of the Fermi arcs and the SPPs was studied in Refs.~\cite{Song:2017,Andolina:2018,Losic:2018,Gorbar-Sukhachov:2019-FAH,Adinehvand-Jafari:2019}. By using semiclassical~\cite{Song:2017} and quantum-mechanical nonlocal~\cite{Andolina:2018} approaches, it was found that the constant frequency contours of the surface plasmons become strongly anisotropic. In addition, as was shown in Refs.~\cite{Gorbar-Sukhachov:2019-FAH,Adinehvand-Jafari:2019}, a gapless Fermi arc collective mode could emerge.

The dispersion relations of surface plasmons can be measured by the scattering-type near-field optical spectroscopy (for a recent review, see Ref.~\cite{Basov-rev:2016}) as well as the momentum-resolved electron energy loss spectroscopy (see, e.g., Ref.~\cite{Wang-Zhang:1995} and references therein). Experimentally, the electron energy loss in Weyl semimetals was recently studied in Ref.~\cite{Chiarello:2018}. The SPPs were experimentally investigated in the type-II Weyl semimetal WTe$_2$ in Ref.~\cite{Tan-Wang:2018}. The nonreciprocity of the SPPs can be used to develop unidirectional optical devices~\cite{Dotsch-Popkov:2005} such as nonreciprocal circulators, nonreciprocal Mach--Zehnder interferometers, one-way optical waveguides~\cite{Takeda:2008}, etc. Tuning the thickness of a Weyl semimetal, dielectric constants of surrounding media, and the direction of the chiral shift provides efficient means to control the strength of the nonreciprocity. However, such a tuning cannot be performed {\it in situ}, which is crucial for creating easily controllable devices. In this study, we propose a different way to control the nonreciprocity of the SPPs connected with the effect of strains in Weyl semimetals.

A remarkable property of mechanical strains in Weyl semimetals is their ability to induce pseudoelectromagnetic fields~\cite{Zhou-Shi:2013,Zubkov:2015,Cortijo-Vozmediano:2015,Cortijo:2016wnf,Grushin-Vishwanath:2016,Pikulin:2016,Liu-Pikulin:2016,Ilan-Pikulin:rev-2019}. Unlike the ordinary electromagnetic fields $\mathbf{E}$ and $\mathbf{B}$, their pseudoelectromagnetic counterparts $\mathbf{E}_5$ and $\mathbf{B}_5$ couple to the left-handed and right-handed particles with opposite signs. A pseudoelectric field $\mathbf{E}_{5}$, for instance, can be created by dynamically stretching or compressing the sample. A nonzero pseudomagnetic field $\mathbf{B}_{5}$ is generated, e.g., by applying a static torsion~\cite{Pikulin:2016,Arjona-Vozmediano:2018} or bending the sample~\cite{Liu-Pikulin:2016}. A typical magnitude of the pseudomagnetic field $B_5$ is estimated to be about $0.3~\mbox{T}$ in the former case and about $15~\mbox{T}$ in the latter case. While dynamical pseudoelectromagnetic fields allow for interesting effects such as the acoustogalvanic effect~\cite{Sukhachov:2019}, for the purposes of this study, it will be sufficient to consider only static deformations in Weyl semimetals with broken TR symmetry. In this model, we found that strains affect the spectrum of the SPPs by reducing their frequencies and even leading to nonreciprocity. Moreover, deformations can be used to tune the localization of the SPPs.

The paper is organized as follows. The model, key notions, and numerical estimates of model parameters are provided in Sec.~\ref{sec:model}. The SSPs for the perpendicular, Faraday, and Voigt configurations of the chiral shift and wave vector are investigated in Sec.~\ref{sec:results}. The obtained results are summarized in Sec.~\ref{sec:Summary}. The effects of a nonuniform chiral shift profile at the surface of Weyl semimetals are discussed in Appendix~\ref{sec:chiral-shift}. Throughout this study, we set $k_{B}=1$.

\section{Model}
\label{sec:model}

\begin{figure}[t]
	\centering
	\subfigure[]{\includegraphics[height=0.2\textwidth]{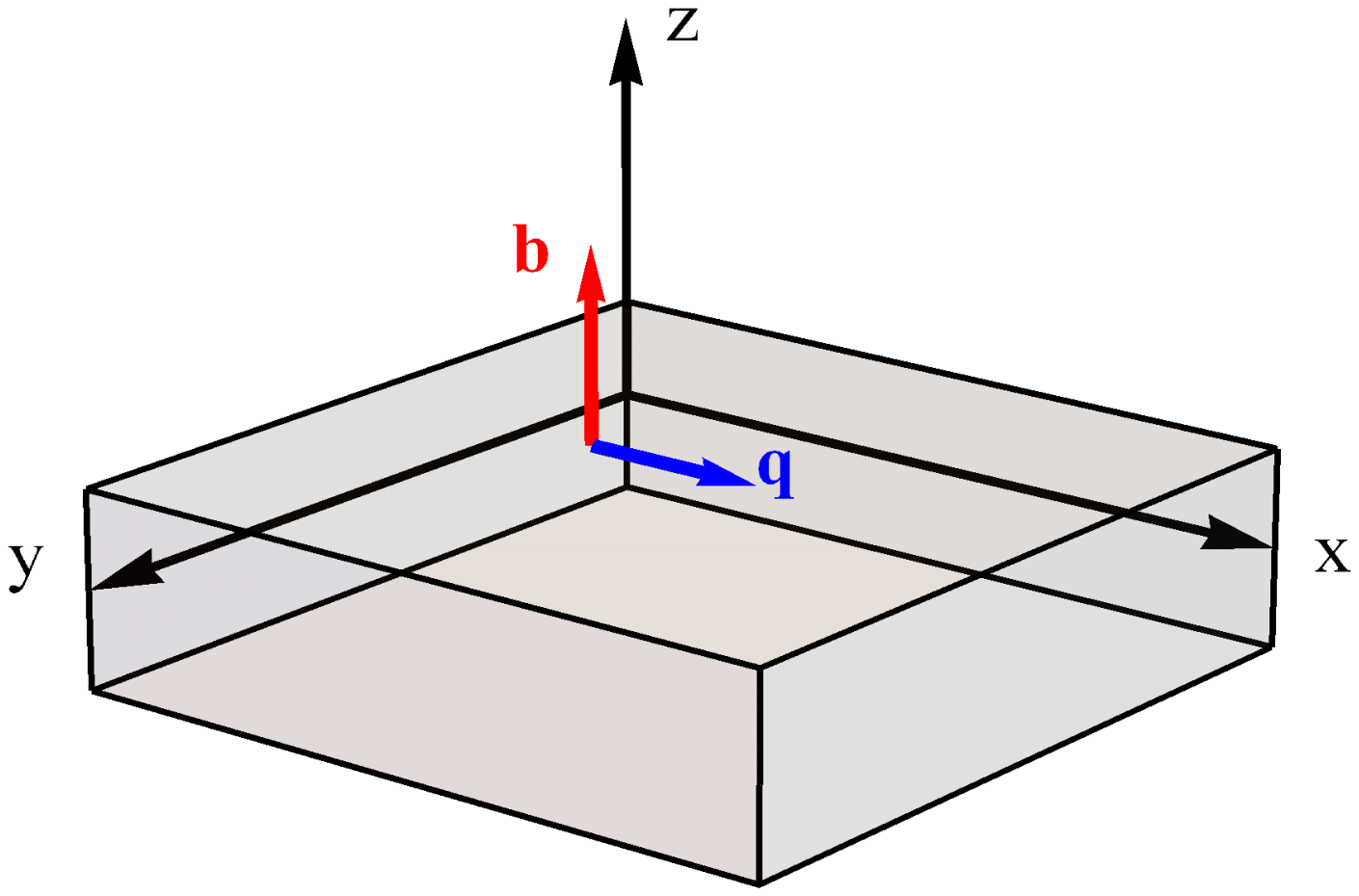}}
	\hspace{0.02\textwidth}
	\subfigure[]{\includegraphics[height=0.2\textwidth]{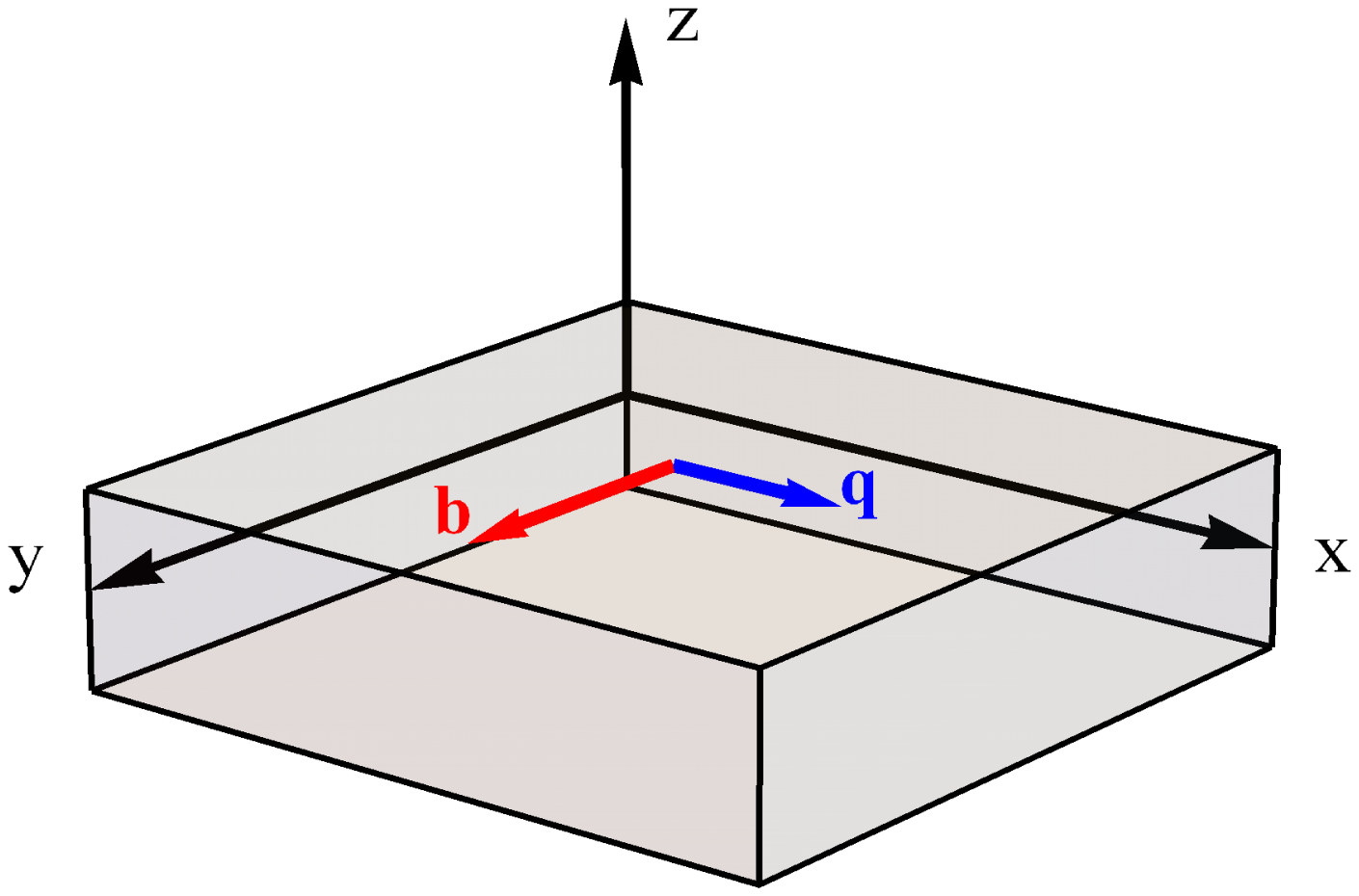}}
	\hspace{0.02\textwidth}
	\subfigure[]{\includegraphics[height=0.2\textwidth]{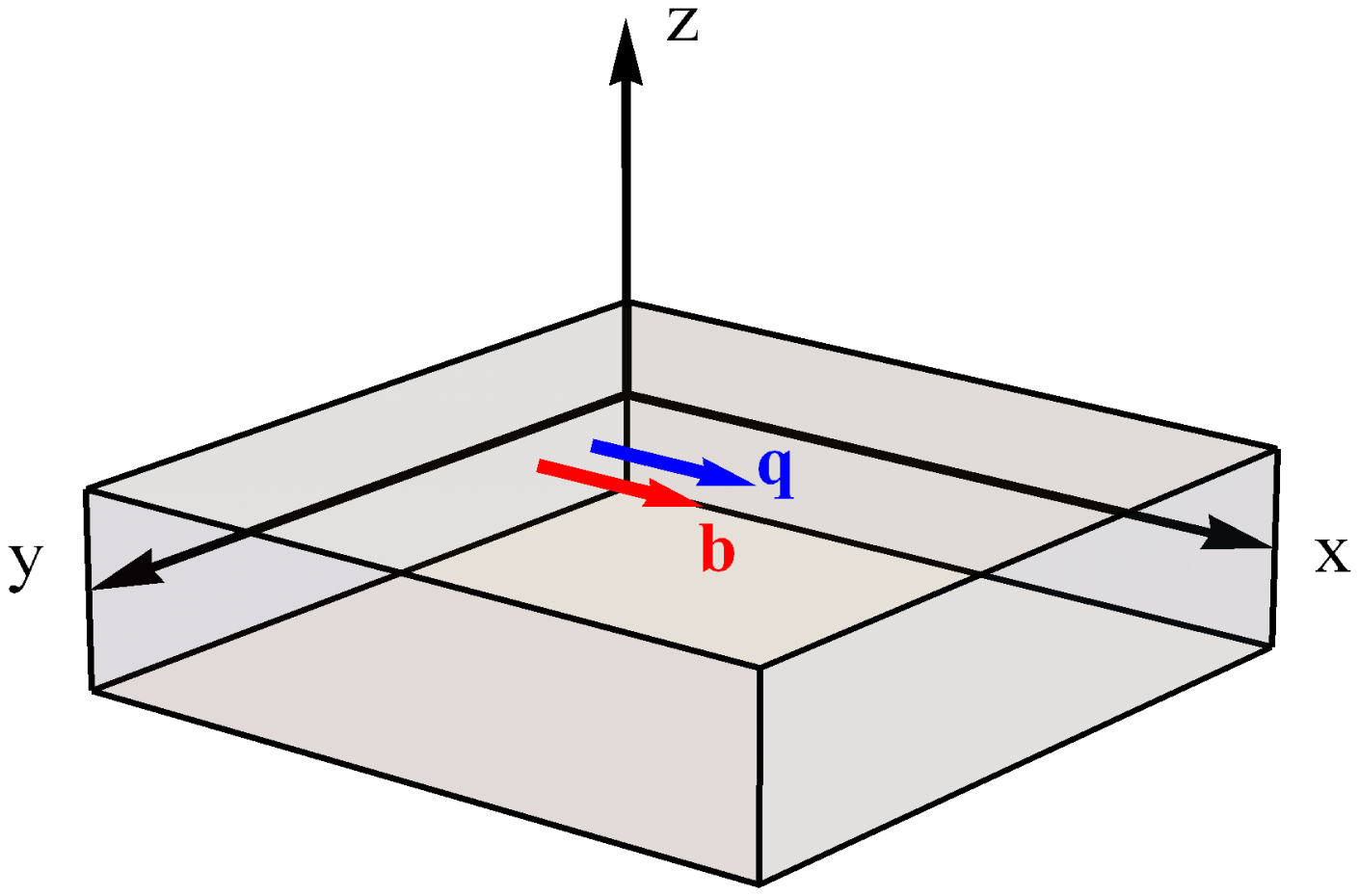}}
	\caption{Schematic setup for the perpendicular configuration $\mathbf{b}\perp\mathbf{q}$ and $\mathbf{b}\parallel\hat{\mathbf{z}}$ (panel (a)), the Voigt configuration  $\mathbf{b}\perp\mathbf{q}$ and $\mathbf{b}\perp\hat{\mathbf{z}}$ (panel (b)), and the Faraday configuration $\mathbf{b}\parallel\mathbf{q}$ and $\mathbf{b}\perp\hat{\mathbf{z}}$ (panel (c)). Here $\mathbf{b}$ is the chiral shift vector, $\mathbf{q}$ is the wave vector of surface plasmons, and $\hat{\mathbf{z}}$ is the unit vector in the $z$ direction. The slab is infinite along the $x$ and $y$ directions and has the width $2d$ in the $z$ direction.
    }
	\label{fig:setup}
\end{figure}

Let us begin with defining the model of a Weyl semimetal and presenting general equations for the SPPs. We assume that the Weyl semimetal has the form of a slab of finite thickness $2d$ along the $z$ direction. The corresponding setup together with three configurations of the chiral shift $\mathbf{b}$ and the wave vector $\mathbf{q}$ of the SPPs is shown in Fig.~\ref{fig:setup}. For the slab of a sufficiently large thickness, the SSPs on its surfaces overlap weakly and can be considered as independent.
In this simplified case, one assumes that the Weyl semimetal is situated at $z > 0$ and vacuum is at $z < 0$. In view of the translational invariance along the interface, the electric field $\mathbf{E}$ is sought as a plane wave with frequency $\omega$ and wave vector along the surface $\mathbf{q}=(q_x,q_y)$, i.e.,
\begin{equation}
\label{model-electric-field-ansatz}
\mathbf{E} \propto e^{-i\omega t + iq_xx+iq_yy}\,e^{-\kappa |z|},
\end{equation}
which decays exponentially away from the boundary for $\kappa > 0$. The field in vacuum is sought in the same form, however, with a different decay constant $\kappa_0$. The electric field is determined by the following equation:
\begin{equation}
\bm{\nabla}\times\left[\bm{\nabla}\times \mathbf{E}\right] =-\frac{1}{c^2}\frac{\partial^2}{\partial t^2}\,\mathbf{D},
\label{model-wave-equation}
\end{equation}
where $\mathbf{D}$ is the displacement electric field and $c$ is the speed of light. The same equation where $\mathbf{D}$ is replaced with $\mathbf{E}$ should be used in vacuum.

\subsection{Hamiltonian and main equations}
\label{sec:model-H}

To demonstrate the effect of strain-induced axial gauge fields on the SPPs in Weyl semimetals, it suffices to consider the minimal model of a Weyl semimetal with a single pair of Weyl nodes separated by $2\mathbf{b}$ in momentum. The corresponding Hamiltonian has the following form:
\begin{equation}
\label{model-H-chi}
H_{\lambda}=-\mu +\lambda \hbar v_F \bm{\sigma} \cdot\left(-i\bm{\nabla} + \lambda \frac{e}{c\hbar} \mathbf{A}_5(\mathbf{r}) -\lambda \mathbf{b}\right).
\end{equation}
Here $\lambda=\pm$ is the chirality of Weyl nodes, $\mu$ is the electric chemical potential, $v_F$ is the Fermi velocity, $\bm{\sigma}$ is the vector of Pauli matrices, and $\mathbf{A}_5(\mathbf{r})$ is the axial gauge field. The latter can be induced by strains~\cite{Zhou-Shi:2013,Zubkov:2015,Cortijo-Vozmediano:2015}. Moreover, a coordinate-dependent axial gauge field appears necessarily at the surface of a Weyl semimetal, where the chiral shift terminates~\cite{Chernodub-Vozmediano:2014,Grushin-Vishwanath:2016,Grushin:2018,Benito-Matias-Gonzalez:2020}. The dependence of $\mathbf{A}_5$ on coordinates and the direction of the chiral shift $\mathbf{b}$ will be specified later in Secs.~\ref{sec:Estimates} and \ref{sec:results}. The effects of nonuniform chiral shift profile are considered in Appendix~\ref{sec:chiral-shift}. In particular, we found that surface collective modes become delocalized when the profile of the chiral shift is sufficiently nonuniform.

In order to determine the displacement electric field $\mathbf{D}$, the dependence of the electric current density $\mathbf{j}$ on the electric field $\mathbf{E}$ should be specified. In addition to the usual Ohm's current, it is well known \cite{Yang-Lu:2011,Burkov-Balents:2011,Burkov-Hook:2011,Grushin:2012,Zyuzin:2012,Goswami:2013,Burkov:2014} that a Weyl semimetal with broken TR symmetry has the AHE current, which is perpendicular to the electric field. This is the origin of the gyrotropic effects observed in Weyl semimetals even in the absence of a magnetic field.

In the model (\ref{model-H-chi}), the AHE current has the form
\begin{equation}
\label{model-current-AHE}
\mathbf{j}_{\text{{\tiny AHE}}} =-\frac{e^2}{2\pi^2 \hbar} \left[\mathbf{b} \times\mathbf{E}\right] +\frac{e^3}{2\pi^2 \hbar^2 c} \left[\mathbf{A}_5 \times\mathbf{E}\right].
\end{equation}
Thus, the explicit expression for the displacement vector $\mathbf{D}$ is
\begin{equation}
\label{model-displacement-field}
\mathbf{D}=\left[\varepsilon(\omega)+\frac{4\pi i}{\omega}\sigma\right] \mathbf{E} - \frac{2ie^2}{\pi \hbar \omega} \left[\mathbf{b} \times \mathbf{E}\right]
+ \frac{2ie^3}{\pi \hbar^2 c \omega} \left[\mathbf{A}_5 \times \mathbf{E}\right],
\end{equation}
where $\sigma$ describes the real part of the electric conductivity related to disorder and $\varepsilon(\omega)$ is the frequency-dependent dielectric constant of Weyl semimetal. For simplicity, we assumed that $\varepsilon(\omega)$ does not depend on the wave vector $\mathbf{q}$. This approximation is justified if the inverse wave vector of SPPs is larger than the inverse Fermi wave vector. Then, the frequency dependence has the standard form $\varepsilon(\omega)=\varepsilon_{\infty}(1-\Omega^2_{\rm e}/\omega^2)$, where $\varepsilon_{\infty}$ is the high-frequency dielectric constant and
\begin{equation}
\label{model-CMP-k=0-Langmuir}
\Omega_{\rm e}^2 = \frac{4e^2}{3\pi\hbar^3 v_F \varepsilon_{\infty}}\left(\mu^2 +\frac{\pi^2 T^2}{3}\right)
\end{equation}
is the plasma or Langmuir frequency. Here $T$ is temperature.

The profiles of electromagnetic fields and frequencies of the corresponding collective modes are determined by solving Eq.~(\ref{model-wave-equation}) with the appropriate boundary conditions. For these conditions, we demand, as usual, the continuity of the parallel components of electric and normal components of magnetic fields. These magnetic fields are generated dynamically by oscillating electric currents and fields. Further, since no external charges and currents are present, the perpendicular components of the displacement field and parallel components of the magnetic field are also continuous. For example, by using ansatz (\ref{model-electric-field-ansatz}), a homogeneous system of linear algebraic equations can be derived in the case of a semi-infinite slab. The zeros of the determinant of this system define the dispersion relation of SPPs. As we will show below, the case of strained Weyl semimetal is more complicated and one can no longer look for solution in form (\ref{model-electric-field-ansatz}).

\subsection{Model parameters}
\label{sec:Estimates}

In order to provide a direct relation to experiments, we quantify the values of model parameters in realistic materials. For definiteness, we use in our analysis the numerical constants valid for the Dirac semimetal Cd$_3$As$_2$~\cite{Freyland-Madelung:book,Wang-Yamazaki:2007,Neupane-Hasan-Cd3As2:2014,Liu-Chen-Cd3As2:2014,Li-Yu-Cd3As2:2015}:
\begin{equation}
\label{estimates-parameters-num}
v_{\rm F}\approx 1.5\times 10^8~{\rm cm/s}, \quad \mu\approx 200~{\rm meV}, \quad b\approx 1.6~{\rm nm}^{-1},
\end{equation}
where the chiral shift is estimated as the distance between two Dirac points in Cd$_3$As$_2$. In addition, the dielectric constant of the Weyl semimetal candidate Eu$_2$Ir$_2$O$_7$, $\varepsilon_{\infty}=13$~\cite{Sushkov-Drew:2015}, is used.

Then, according to Eq.~(\ref{model-CMP-k=0-Langmuir}), the plasma frequency at $T\to0$ can be estimated as
\begin{equation}
\label{estimates-Omegae}
\Omega_{\rm e} \approx 6.6\times10^{13}~\mbox{s}^{-1}.
\end{equation}
This frequency corresponds to the following characteristic length scale:
\begin{equation}
\label{estimates-char-length}
\frac{c}{\Omega_e} \approx 4.5~\mu\mbox{m}.
\end{equation}
Note that the thickness of films of  Weyl and Dirac semimetal could be even smaller than the characteristic length scale.
For example, films of the Dirac semimetal Cd$_3$As$_2$ with the thickness $2d\approx 35-100~\mbox{nm}$~\cite{Schumann-Stemmer-Cd3As2:2019,Nishihaya-Kawasaki-Cd3As2:2019} and the Weyl semimetals NbP and TaP with the thickness $2d\approx 9-70~\mbox{nm}$~\cite{Bedoya-Pinto-Parkin-NbP:2020} can be grown.
The characteristic frequency corresponding to the Weyl node separation is given by
\begin{equation}
\label{estimates-omegab}
\omega_b=\frac{2 e^2 b}{\pi \hbar \varepsilon_{\infty}}\approx 1.7\times10^{14}~\mbox{s}^{-1}\approx2.6\,\Omega_e.
\end{equation}
It is interesting to note that this frequency is comparable to $\Omega_e$. This suggests that the effects related to the Weyl nodes separation could be indeed significant in real materials.

Further, let us provide estimates of strain magnitude. We start with the case of bending about the $y$ axis. The corresponding components of the displacement field $\mathbf{u}$ are~\cite{Landau:t7}
\begin{eqnarray}
\label{estimates-u-bend}
u_x=\frac{u_0}{d} xz, \quad u_z=-\frac{u_0}{2d}\left(x^2+D_{\rm L}z^2\right).
\end{eqnarray}
Here $u_0$ is the maximum stress and $D_{\rm L}$ is a certain function of the Lam\'{e} coefficients. The corresponding strain-induced axial gauge field for $\mathbf{b}\parallel\hat{\mathbf{x}}$ can be estimated as~\cite{Cortijo-Vozmediano:2015}
\begin{equation}
\label{estimates-A5x-bend}
A_{5,x} \simeq -\frac{c\hbar}{e} \beta_G b_x u_{xx} = -\frac{c\hbar \beta_G b_x u_0}{ed} z,
\end{equation}
where $\hat{\mathbf{x}}$ is the unit vector in the $x$ direction, $\beta_{G}\simeq1$ is the Gr\"{u}neisen parameter, and the standard definition of the strain tensor was used, $u_{ij}=\left(\partial_iu_j+\partial_j u_i\right)/2$. Then, the effective axial field strength, which is defined as
\begin{equation}
\label{estimates-tA5-def}
\tilde{A}_{5} \simeq |A_5| \frac{d}{z},
\end{equation}
reads as
\begin{equation}
\label{estimates-tA5x-bend}
\tilde{A}_{5,x} \simeq  \frac{c\hbar \beta_G b_x u_0}{e}.
\end{equation}

We find it convenient to quantify the magnitude of strain by the following dimensionless parameter:
\begin{eqnarray}
\label{estimates-beta-def}
\beta = \sqrt{\frac{c\omega_b}{\Omega_e^2 b l^2}},
\end{eqnarray}
where $l^2 = \hbar c d/(e\tilde{A}_5)$. In the case of bending, it is estimated as
\begin{eqnarray}
\label{estimates-beta-bend}
\beta = \sqrt{\frac{c\omega_b}{\Omega_e^2 b l^2}} = \sqrt{\frac{2e^3 \tilde{A}_{5,x}}{\pi \hbar^2 \varepsilon_{\infty} \Omega_e^2 d}} \simeq \sqrt{\frac{2u_0e^2 c \beta_G b_x}{\pi \hbar \varepsilon_{\infty} \Omega_e^2 d}} \approx 1.6 \sqrt{\frac{c}{\Omega_e d} u_0}.
\end{eqnarray}
As expected, the strain effects are well manifested in sufficiently thin films. For example, even for $u_0=1\%$ and $d=0.1c/\Omega_e$, the dimensionless parameter $\beta\approx0.5$. In such a case, however, the SPPs on the opposite surfaces hybridize notably.

In the case of an inhomogeneous stretching along the $z$ direction, the $z$ component of the displacement vector is
\begin{eqnarray}
\label{estimates-u-stretch}
u_z =z \frac{f(z)}{2d} =z^2 \frac{f(d)-f(-d)}{(2d)^2},
\end{eqnarray}
where we assumed a linear dependence of the function $f(z)$ on coordinates. Then
\begin{equation}
\label{estimates-A5z-stretch}
A_{5,z} \simeq -\frac{c\hbar}{e} \beta_G b_z u_{zz} = -\frac{c\hbar}{e} \beta_G b_z z \frac{f(d)-f(-d)}{2d^2}.
\end{equation}
The corresponding effective axial field strength and the dimensionless parameter $\beta$ are
\begin{equation}
\label{estimates-tA5z-stretch}
\tilde{A}_{5,z} \simeq  \frac{c\hbar}{e} \beta_G b_z \frac{\left|f(d)-f(-d)\right|}{2d}
\end{equation}
and
\begin{eqnarray}
\label{estimates-beta-stretch}
\beta \simeq \sqrt{\frac{2ce^2}{\pi \hbar \varepsilon_{\infty} \Omega_e^2 d} \beta_G b_z \frac{\left|f(d)-f(-d)\right|}{2d}} \approx 1.6 \sqrt{\frac{c}{\Omega_e d} \frac{\left|f(d)-f(-d)\right|}{2d}},
\end{eqnarray}
respectively. As in the case of bending, the relative deformation $\left|f(d)-f(-d)\right|/(2d)$ could reach a few percents.

\section{Results for surface plasmon polaritons}
\label{sec:results}

In this section, we discuss the results for the dispersion relations of SPPs in Weyl and Dirac semimetals and show how strains affect them.
Let us consider first the case of a Dirac semimetal with $\mathbf{b}=\mathbf{A}_5=\mathbf{0}$. Then it is easy to obtain that the dispersion of the SPPs coincides with that in ordinary metals~\cite{Ritchie-Wilems:1969,Barton:rev-1979,Boardman:book,Pitarke-Echenique:rev-2006} and is determined by the following relation:
\begin{equation}
\varepsilon_1\kappa_0+ \kappa=0,
\label{results-Dirac-semimetal}
\end{equation}
where $\kappa=\sqrt{q^2-\varepsilon_1\omega^2/c^2}$ and $\kappa_0=\sqrt{q^2-\omega^2/c^2}$. The AHE currents and the corrections due to the axial fields generated by strains cancel out for Dirac semimetals.

Let us present now the results for Weyl semimetals with broken $\mathcal{T}$ symmetry ($\mathbf{b} \neq \mathbf{0}$). As we will see below and as was noted in, e.g., Ref.~\cite{Hofmann-DasSarma:2016}, the SPPs in Weyl semimetals resemble the magnetoplasmons in conventional metals~\cite{Chiu-Quinn:1972,Wallis-Hartstein:1974,Kushwaha-Halevi:1987,Boardman:book}.

It is convenient to rewrite Eq. (\ref{model-wave-equation}) as
\begin{equation}
\label{results-wave-eq}
\nabla(\nabla \cdot \mathbf{E}) - \Delta \mathbf{E} = \frac{\omega^2}{c^2}\left(\varepsilon_1 \mathbf{E} - i \varepsilon_2 [\hat{\mathbf{b}}\times \mathbf{E}] + i \varepsilon_2 \frac{z}{b l^2}[\hat{\mathbf{A}}_5\times \mathbf{E}]\right),
\end{equation}
where $\varepsilon_{1} = \varepsilon(\omega) +4\pi i \sigma/\omega$, $\varepsilon_2 = \varepsilon_{\infty}\omega_b/\omega$, and $\hat{\mathbf{A}}_5$ is the unit vector in the direction of $\mathbf{A}_5$. A nonzero conductivity $\sigma$ leads to a dissipation of the SPPs. For the sake of simplicity, we will ignore it in the rest of the study.

The explicit form of Eq.~(\ref{results-wave-eq}) is
\begin{equation}
\label{results-wave-eq-expl}
\begin{pmatrix}
q_y^2 - \partial_z^2 & -q_xq_y & i q_x \partial_z \\
-q_x q_y & q_x^2 - \partial_z^2 & i q_y \partial_z \\
i q_x \partial_z & i q_y \partial_z & q_x^2 + q_y^2
\end{pmatrix}
\begin{pmatrix}
E_x \\
E_y \\
E_z
\end{pmatrix}
=
\frac{\omega^2}{c^2}
\begin{pmatrix}
\varepsilon_1& i \hat{b}_z \varepsilon_2 -\frac{i z}{b l^2} \hat{A}_{5,z} \varepsilon_2& -i \hat{b}_y \varepsilon_2 + \frac{i z}{b l^2} \hat{A}_{5,y} \varepsilon_2 \\
-i \hat{b}_z \varepsilon_2 +\frac{i z}{b l^2} \hat{A}_{5,z} \varepsilon_2& \varepsilon_1&  i \hat{b}_x \varepsilon_2 -  \frac{i z}{b l^2} \hat{A}_{5,x} \varepsilon_2\\
i\hat{b}_y \varepsilon_2 - \frac{i z}{b l^2} \hat{A}_{5,y} \varepsilon_2& -i \hat{b}_x \varepsilon_2  + \frac{i z}{b l^2} \hat{A}_{5,x} \varepsilon_2 & \varepsilon_1\\
\end{pmatrix}
\begin{pmatrix}
E_x \\
E_y \\
E_z
\end{pmatrix}.
\end{equation}

It is easy to check that the electric field $\mathbf{E}$ takes the following form in vacuum:
\begin{equation}
\mathbf{E}_0= \left(E_x(\pm d), E_y(\pm d), \pm i\frac{E_x(\pm d) q_x + E_y(\pm d) q_y}{\kappa_0}\right) e^{i q_x x + i q_y y - \kappa_0 |z\mp d| - i \omega t}.
\end{equation}
Here $\pm$ corresponds to the upper ($+$) and lower ($-$) vacuum half-spaces.

\subsection{Perpendicular configuration}
\label{sec:results-sol-perpendicular}

Let us start our analysis of the SPPs in strained Weyl semimetals with the perpendicular configuration $\mathbf{b}\perp\mathbf{q}$ and $\mathbf{b}\parallel\hat{\mathbf{z}}$ (see Fig.~\ref{fig:setup}(a)). Without the loss of generality, we set the wave vector of the SPPs pointing in the $x$ direction, i.e., $\mathbf{q}\parallel\hat{\mathbf{x}}$. Further, we assume that $\hat{\mathbf{A}}_5\parallel \hat{\mathbf{z}}$. As was discussed in Sec.~\ref{sec:Estimates}, this axial gauge field could be generated by stretching the sample inhomogeneously along the $z$ axis with $u_z\propto z f(z)/d$ and $f(z)=z$.

It is straightforward to show that the matrix equation (\ref{results-wave-eq-expl}) can be rewritten as a fourth-order ordinary differential equation
\begin{eqnarray}
\label{results-sol-Perp-eq}
&&\frac{\varepsilon_1}{\varepsilon_2 \kappa^2\left[1-z/(bl^2)\right]}E_y^{(4)} +\frac{2\varepsilon_1}{\varepsilon_2 bl^2 \kappa^2 \left[1-z/(bl^2)\right]^2} E_y^{(3)} +\frac{2\varepsilon_1}{\varepsilon_2 (bl^2)^2\left[1-z/(bl^2)\right]^3}\left[\frac{1}{\kappa^2} -(bl^2)^2\left(1-\frac{z}{bl^2}\right)^{2}\right] E_y^{\prime \prime} \nonumber\\
&&-\frac{2\varepsilon_1}{\varepsilon_2 bl^2 \left[1-z/(bl^2)\right]^2} E_y^{\prime}-\frac{1}{\varepsilon_2 \left[1-z/(bl^2)\right]^3} \Bigg[\frac{2\varepsilon_1}{(bl^2)^2} -q^2 \varepsilon_1 \left(1-\frac{z}{bl^2}\right)^2 +\frac{\omega^2 \varepsilon_1^2}{c^2} \left(1-\frac{z}{bl^2}\right)^2 \nonumber\\
&&-\frac{\omega^2 \varepsilon_2^2}{c^2} \left(1-\frac{z}{bl^2}\right)^4\Bigg] E_y=0,
\end{eqnarray}
where we used
\begin{eqnarray}
E_x &=& -ic^2\frac{E_y^{\prime \prime} -\kappa^2 E_y}{\varepsilon_2\omega^2 \left[1-z/(bl^2)\right]},\\
E_z &=& -\frac{iq E_x^{\prime}}{\kappa^2},
\end{eqnarray}
and $q=q_x$. Note that since both $E_x$ and $E_z$ are generically nonzero, SPPs are not purely longitudinal or transverse waves.

One can check that Eq.~(\ref{results-sol-Perp-eq}) reproduces the results obtained in Ref.~\cite{Hofmann-DasSarma:2016} in the limit of semi-infinite slab $d\to\infty$ and vanishing pseudomagnetic field $l\to\infty$. In particular, the decay constant in ansatz (\ref{model-electric-field-ansatz}) equals
\begin{equation}
\label{results-sol-Perp-semiinf}
\kappa_{\rm P}^2 = \kappa^2  \pm \frac{|\omega \kappa \varepsilon_2|}{c \sqrt{-\varepsilon_1}}.
\end{equation}

The dispersion relation of the SPPs in the finite slab is obtained by solving Eq.~(\ref{results-sol-Perp-eq}) and requiring the continuity of the tangential components of the dynamical magnetic field. The latter condition is equivalent to the continuity of $\partial_z E_y$ and $\partial_z E_x - i q E_z$ at the boundaries. For fields outside the slab, we have $\partial_z E_x - i q E_z = \sign{z} \omega^2E_x/(c^2\kappa_0)$. Therefore, since the tangential components of the electric field are continuous, we obtain
\begin{eqnarray}
\label{results-sol-Perp-char-eq}
\left.\kappa_0 \varepsilon_1 E_x^{\prime} + \frac{z q \kappa_0 \varepsilon_2}{b l^2}E_x +\sign{z}\kappa^2E_x\right|_{z = \pm d} = 0,\\
\label{results-sol-Perp-char-eq-Ey}
\left.\partial_z E_y +\kappa_0\sign{z} E_y\right|_{z = \pm d}=0.
\end{eqnarray}

The case of a finite slab with nonzero $\mathbf{A_5}$ is more complicated. Therefore, we focus on numerical solutions. It is worth noting, however, that analytical analysis could be still performed in the case of short and long wavelengths or, equivalently, $q\to\pm\infty$ and $q\to0$, respectively.
In the latter case, since SPPs are gapless collective modes, $\varepsilon_1\to\infty$ at $\omega\to0$ leading to the divergence of the first term in Eq.~(\ref{results-sol-Perp-char-eq}). Therefore, in order to satisfy the characteristic equation, one should set $\kappa_0 = 0$. This leads to the following dispersion relation at small momenta:
\begin{equation}
\label{results-sol-Perp-qto0}
\omega(q\to0) =c q,
\end{equation}
which is nothing else as the dispersion of light. Thus, neither chiral shift nor strains affect the SPPs at small $q$.

Further, let us consider the short wavelength limit $q\to\pm\infty$. In this case, Eq.~(\ref{results-sol-Perp-eq}) simplifies and can be rewritten as
\begin{equation}
\label{results-sol-Perp-eq-q-inf}
\varepsilon_1 E_y^{(4)} +\frac{2\varepsilon_1}{bl^2\left[1-z/(bl^2)\right]} E_y^{(3)} -2\varepsilon_1 q^2 E_y^{\prime \prime} -\frac{2q^2 \varepsilon_1}{bl^2\left[1-z/(bl^2)\right]} E_y^{\prime}+\varepsilon_1 q^4E_y=0.
\end{equation}
Its general solution is
\begin{equation}
\label{results-sol-Perp-sol-q-inf}
E_y=C_1e^{qz} +C_2e^{-qz} +C_3 z\left[3+ qbl^2\left(2-\frac{z}{bl^2}\right)\right] e^{qz}+C_4 z\left[3- qbl^2\left(2-\frac{z}{bl^2}\right)\right]e^{-qz},
\end{equation}
where $C_i$ with $i=\overline{1,4}$ are constants determined from the boundary conditions (\ref{results-sol-Perp-char-eq}) and (\ref{results-sol-Perp-char-eq-Ey}). By substituting solution (\ref{results-sol-Perp-sol-q-inf}) into Eqs.~(\ref{results-sol-Perp-char-eq}) and (\ref{results-sol-Perp-char-eq-Ey}), we find
\begin{equation}
\label{results-sol-Perp-omega-q-inf}
\omega\left(q \to \pm\infty\right) = \Omega_e \sqrt{\frac{\varepsilon_{\infty}}{1+\varepsilon_{\infty}}}.
\end{equation}
This result agrees with that for conventional surface plasmons~\cite{Ritchie:1957}. It is clear that strain does not induce nonreciprocity in this case.

\begin{figure}[t]
	\centering
	\subfigure[]{\includegraphics[height=0.275\textwidth]{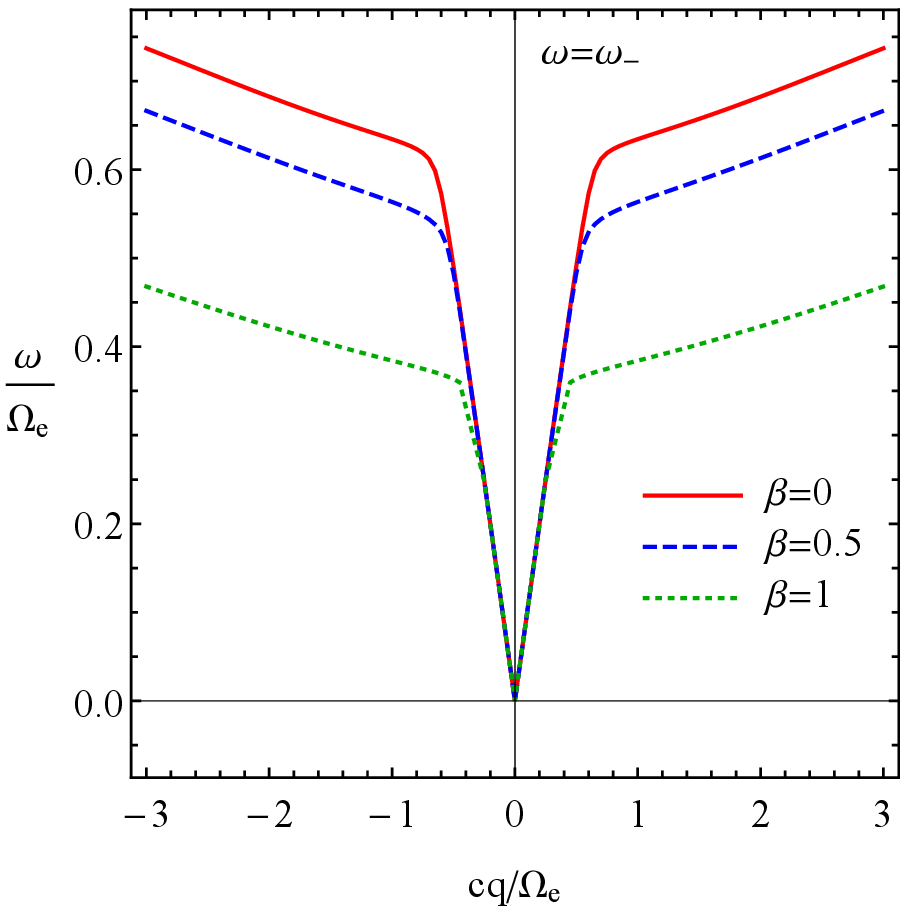}}
	\hspace{0.05\textwidth}
	\subfigure[]{\includegraphics[height=0.275\textwidth]{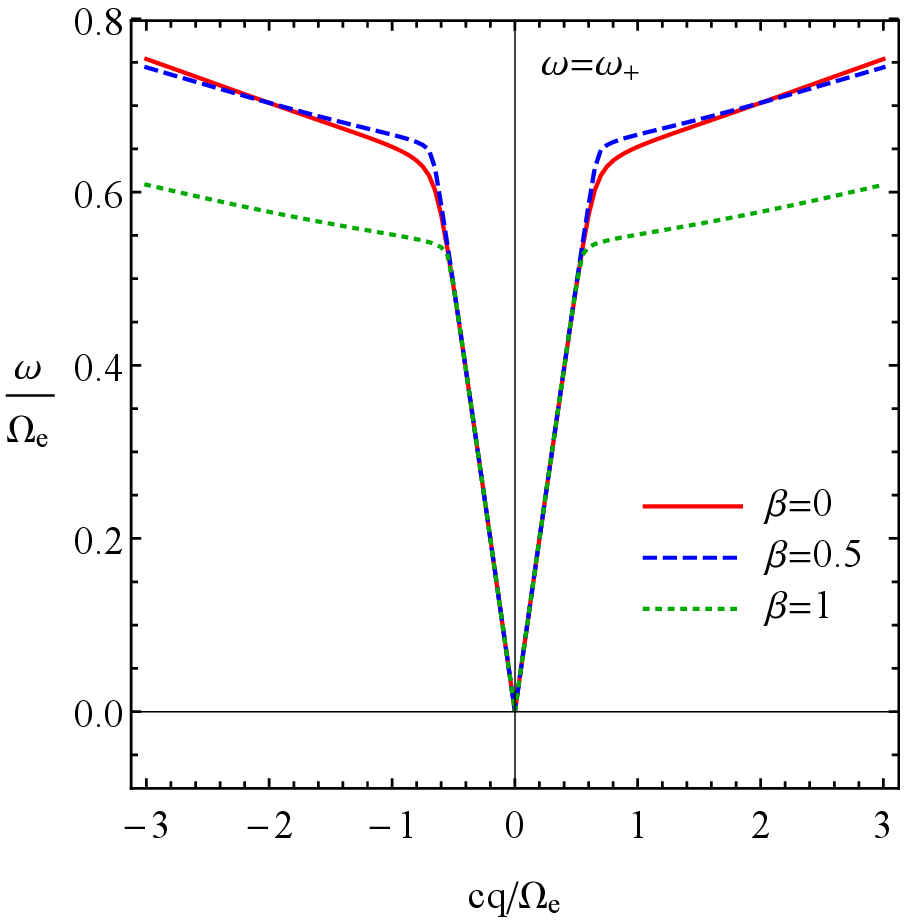}}
	\hspace{0.05\textwidth}
	\subfigure[]{\includegraphics[height=0.275\textwidth]{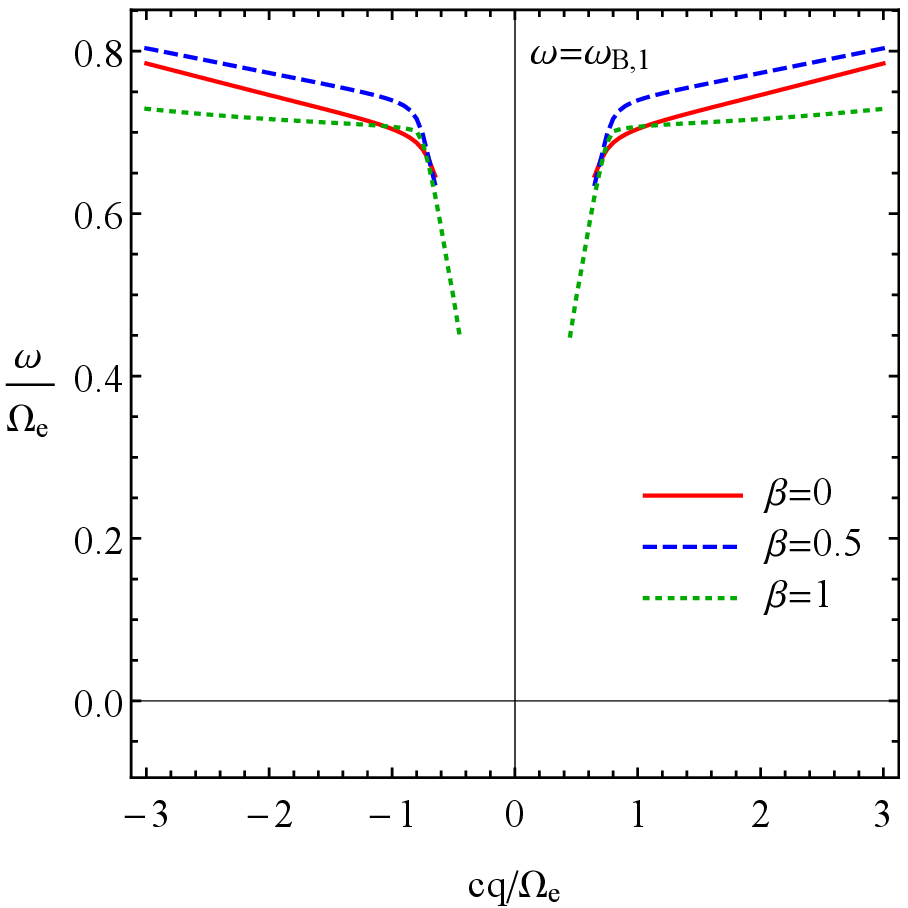}}
	\caption{The dispersion relation of the surface plasmon polaritons in a slab of Weyl semimetal for the perpendicular configuration at $\beta=0$ (red solid lines), $\beta=0.5$ (blue dashed lines), and $\beta=1$ (green dotted lines). Panels (a), (b), and (c) correspond to two SPP branches $\omega_{-}$ and $\omega_{+}$, as well as the lowest bulk mode $\omega_{\rm B,1}$, respectively. We set $d=2c/\Omega_e$ and $\omega_b=\Omega_e$.
    }
	\label{fig:results-perp-omega-few-beta}
\end{figure}

\begin{figure}[t]
	\centering
	\subfigure[]{\includegraphics[height=0.35\textwidth]{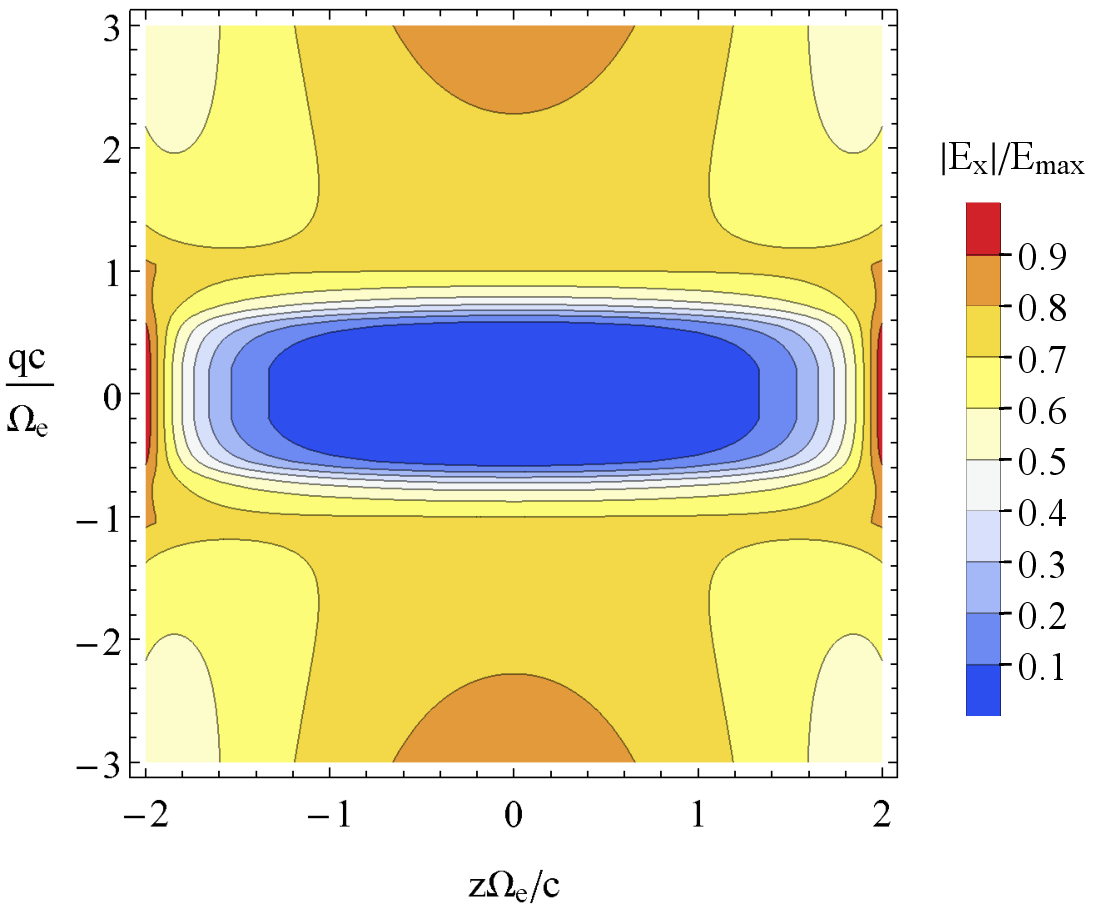}}
	\hspace{0.02\textwidth}
	\subfigure[]{\includegraphics[height=0.35\textwidth]{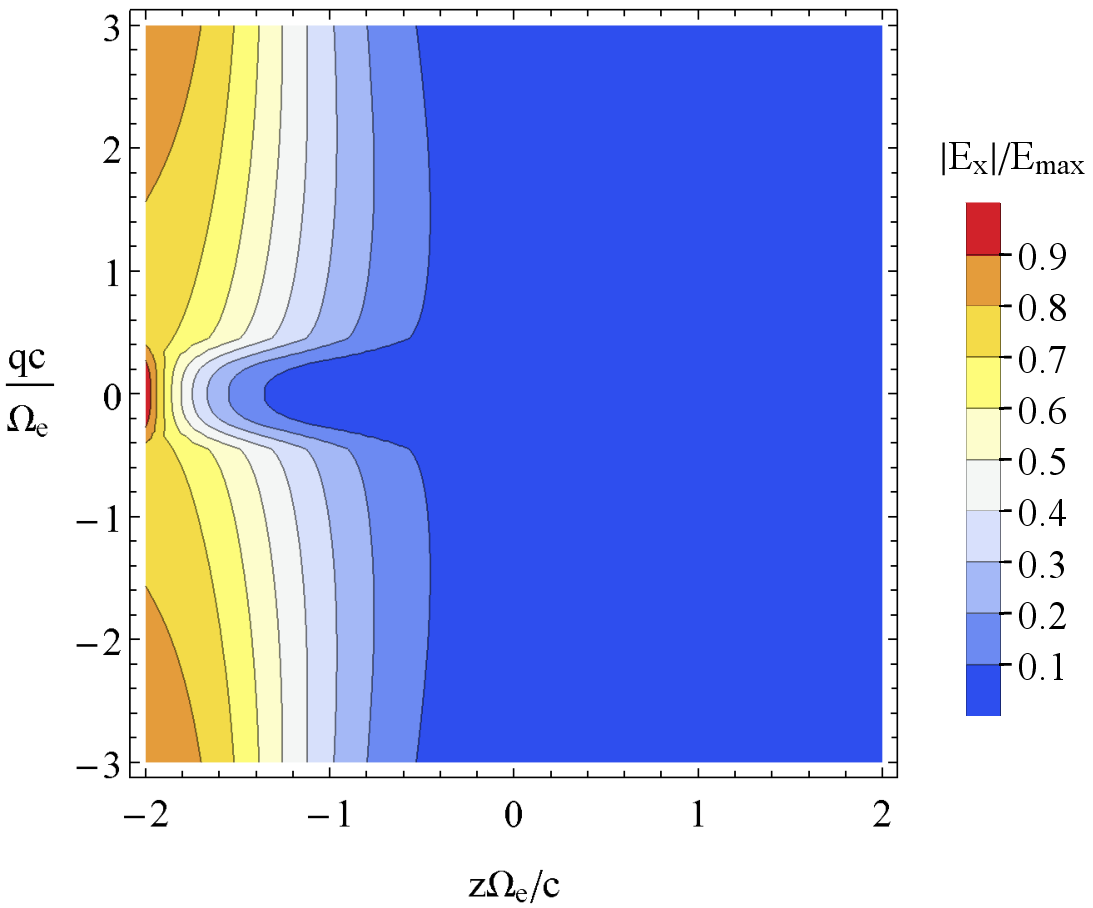}}
	\hspace{0.02\textwidth}
	\subfigure[]{\includegraphics[height=0.35\textwidth]{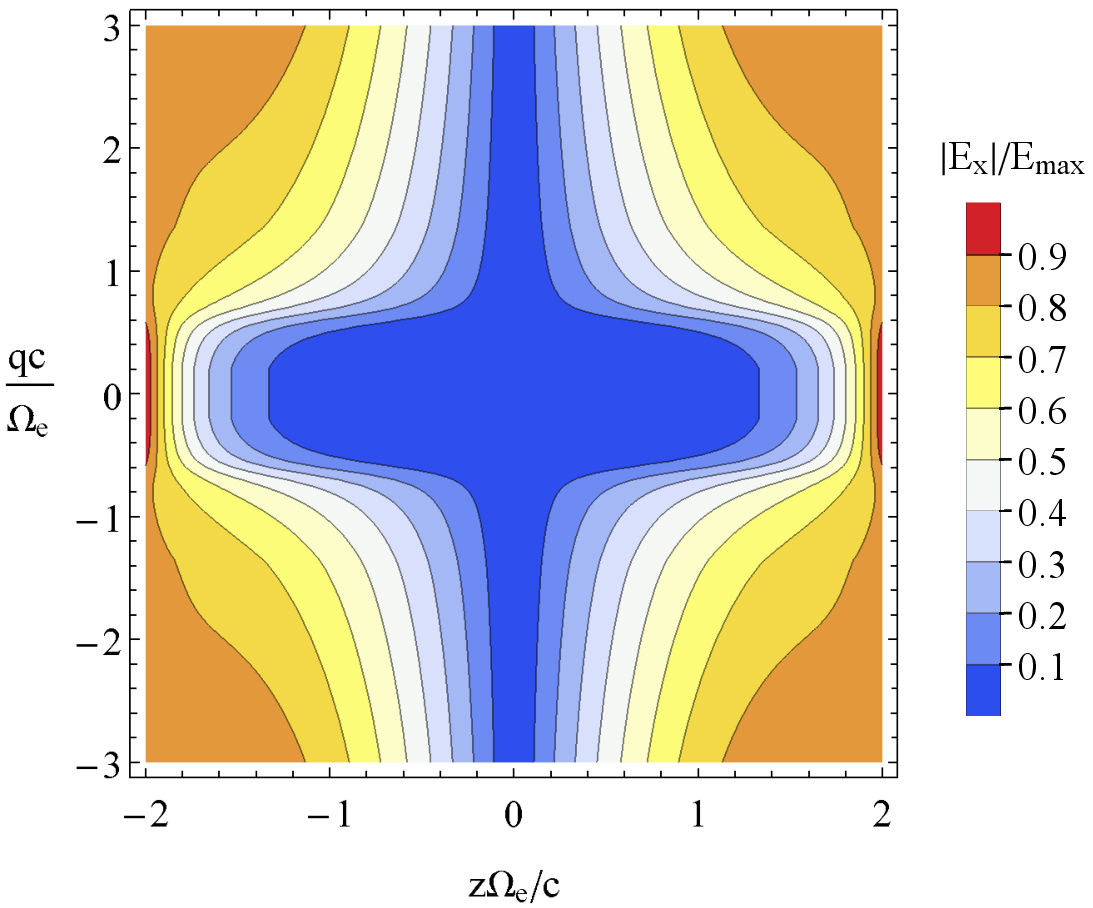}}
	\hspace{0.02\textwidth}
	\subfigure[]{\includegraphics[height=0.35\textwidth]{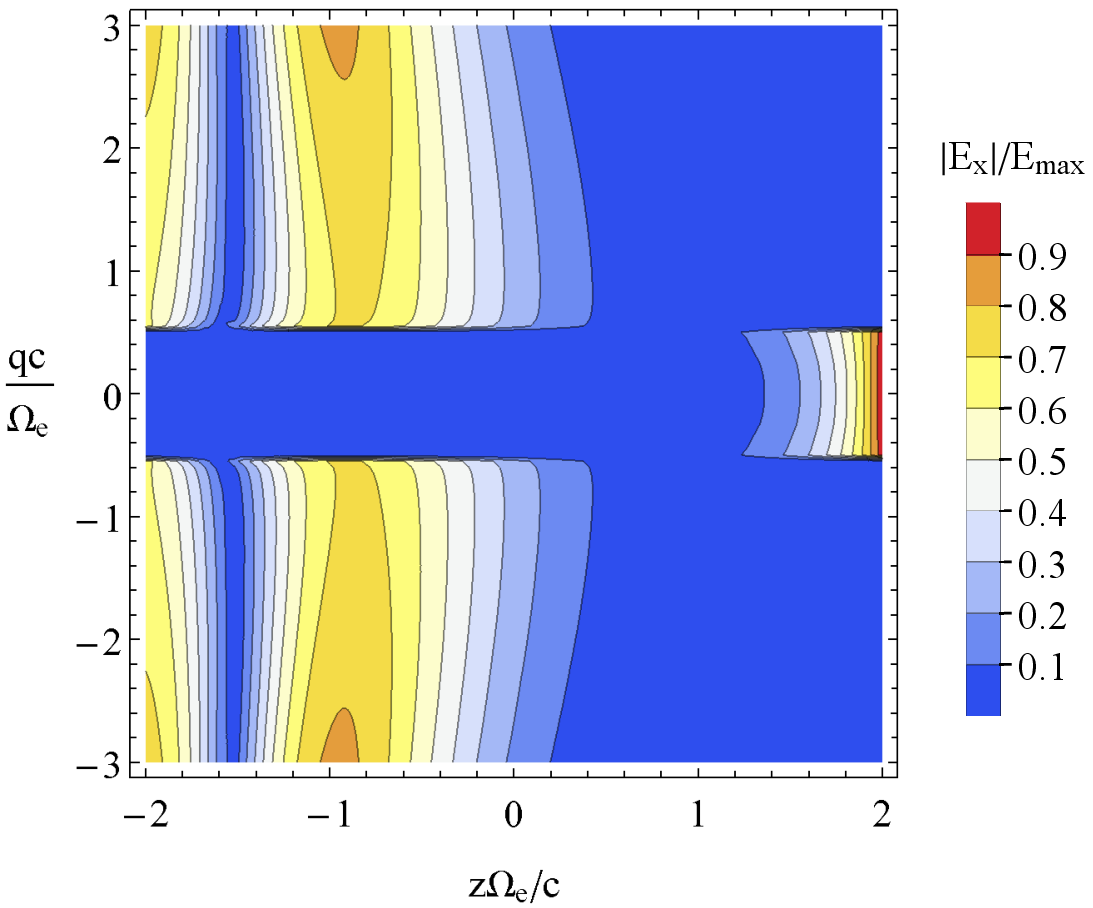}}
	\caption{Profiles of the $x$ component of the electric field $E_x$ in the perpendicular configuration. Top and bottom panels correspond to two SPP branches $\omega_{-}$ and $\omega_{+}$, respectively. Strain strength is $\beta=0$ in panels (a) and (c) as well as $\beta=1$ in panels (b) and (d). We set $d = 2c/\Omega_{e}$ and $\omega_b = \Omega_e$.}
	\label{fig:results-perp-fields}
\end{figure}

The numerical solutions for dispersion relations obtained from Eq.~(\ref{results-sol-Perp-eq}) with the boundary conditions (\ref{results-sol-Perp-char-eq}) and (\ref{results-sol-Perp-char-eq-Ey}) are shown in Fig.~\ref{fig:results-perp-omega-few-beta}, where the case $\beta=0$ corresponds to the absence of strain.
The frequencies $\omega_{+}$ and $\omega_{-}$ correspond to two branches of the SPP spectrum. If the width of the slab is sufficiently large, then these modes can be understood as a combination of the SPPs localized at the opposite surfaces. They are hybridized, however, in a thin slab. Nevertheless, we can still distinguish them by using the symmetry properties of the field component $E_x$ in the unstrained limit. In this case, $\omega_{+}$ and $\omega_{-}$ correspond to the modes with antisymmetric and symmetric distributions of the field, respectively.
Clearly, strain decreases frequencies of the SPPs for intermediate values of $q$. In agreement with the analytical result (\ref{results-sol-Perp-omega-q-inf}), there is no dependence on strain at $q\to\pm \infty$, however. In addition to the SPPs, we also present one of the bulk modes in Fig.~\ref{fig:results-perp-omega-few-beta}(c), which is determined as the lowest delocalized solution.
The field profiles of the SPPs are shown in Fig.~\ref{fig:results-perp-fields}. Unlike the case of semi-infinite slab, where the electric field for the surface modes is localized at the boundary, the field could be relatively large inside a slab of small thickness. The localization becomes more pronounced as the slab width increases. Furthermore, we found that the strain enhances the localization of the SPPs. Depending on its direction, the modes become localized on either top or bottom surface. Therefore, deformations can be used to effectively tune the localization of the SPPs in Weyl semimetals.

\subsection{Voigt configuration}
\label{sec:results-sol-Voigt}

Let us proceed to the Voigt configuration, which is schematically shown in Fig.~\ref{fig:setup}(b). For the sake of definiteness, we set $\mathbf{q}\parallel \hat{\mathbf{x}}$ and $\mathbf{b}\parallel\hat{\mathbf{y}}$.
Further, we assume that $\hat{\mathbf{A}}_5\parallel\hat{\mathbf{y}}$. As we discussed in Sec.~\ref{sec:Estimates}, this axial gauge field can be generated by bending about the $x$ axis producing $\mathbf{A}_5 \propto u_0b_yz \hat{\mathbf{y}}/d$.

Equation~(\ref{results-wave-eq-expl}) takes the following form in the case of the Voigt configuration:
\begin{eqnarray}
\label{results-sol-Voigt-eq}
\varepsilon_1 E_x^{\prime \prime} -E_x \left[\varepsilon_1 q^2 -\frac{q\varepsilon_2}{bl^2} - \frac{\varepsilon_1^2\omega^2}{c^2} + \left(1-\frac{z}{bl^2}\right)^2\frac{\omega^2 \varepsilon_2^2}{c^2} \right]=0.
\end{eqnarray}
Note that the $y$ component of the field is decoupled and does not correspond to plasmon modes. Furthermore, it can be shown that it vanishes after matching with solutions in vacuum.

The $z$ component of the electric field is related to $E_x$ according to
\begin{equation}
\label{results-sol-Voigt-Ez-Ex}
E_z = -i\frac{q}{\kappa^2} E_x^{\prime} + i\frac{\omega^2}{c^2 \kappa^2}\varepsilon_2\left(1 - \frac{z}{b l^2}\right) E_x.
\end{equation}

Let us check that we reproduce the results obtained in the literature if strains are ignored. By using ansatz (\ref{model-electric-field-ansatz}) and taking the limit $l \rightarrow \infty$, the following decay constant is obtained:
\begin{equation}
\label{results-sol-Voigt-kappaV}
\kappa_{V}^2 = q^2 + \frac{\omega^2}{c^2}\left(\frac{\varepsilon_2^2}{\varepsilon_1} - \varepsilon_1\right),
\end{equation}
which agrees with the result in Ref.~\cite{Hofmann-DasSarma:2016}.

In general, Eq.~(\ref{results-sol-Voigt-eq}) should be solved numerically. By requiring the continuity of the tangential component of the magnetic field, which is equivalent to the continuity of $\partial_z E_x - i q E_z$, the following characteristic equation is derived:
\begin{equation}
\label{results-sol-Voigt-char-eq}
\left.\kappa_0 \varepsilon_1 E_x^{\prime} - \kappa_0 \varepsilon_2 q\left(1 - \frac{z}{b l^2}\right)E_x + \sign{z}\kappa^2 E_x \right|_{z = \pm d} = 0.
\end{equation}
Here, the last term stems from the vacuum solution.

Before presenting numerical results, let us investigate the limit of long and short wavelengths, i.e., $q\to0$ and $q\to \pm \infty$, respectively. In the case $q\to0$, the same simple result as in the perpendicular configuration can be obtained [see Eq.~(\ref{results-sol-Perp-qto0})]. For short wavelengths ($q\to \pm\infty$), a solution to Eq.~(\ref{results-sol-Voigt-eq}) can be sought as $E_x(z) = C_1 e^{q z} + C_2 e^{-q z}$. Then, by using Eq.~(\ref{results-sol-Voigt-char-eq}) and retaining only the leading in $1/q$ terms, we obtain
\begin{eqnarray}
\label{results-sol-Voigt-omega-q-inft-plus}
\omega_{\pm}(q\to\infty) = -\omega_b\varepsilon_{\infty}\frac{d\mp bl^2}{2bl^2 (1+\varepsilon_{\infty})} + \frac{\sqrt{(\omega_b\varepsilon_{\infty})^2\left(d\mp bl^2\right)^2 +4b^2l^4 \Omega_e^2 \varepsilon_{\infty}(1+\varepsilon_{\infty})}}{2bl^2(1+\varepsilon_{\infty})},\\
\label{results-sol-Voigt-omega-q-inft-minus}
\omega_{\pm}(q\to-\infty) = \omega_b\varepsilon_{\infty}\frac{d\pm bl^2}{2bl^2 (1+\varepsilon_{\infty})} + \frac{\sqrt{(\omega_b\varepsilon_{\infty})^2\left(d\pm bl^2\right)^2 +4b^2l^4 \Omega_e^2 \varepsilon_{\infty}(1+\varepsilon_{\infty})}}{2bl^2(1+\varepsilon_{\infty})},
\end{eqnarray}
where subscript $\pm$ corresponds to the second ($+$) and first ($-$) branches of the SPP spectrum. As one can see, the spectrum is nonreciprocal. The magnitude of the nonreciprocity for a weak strain and a small chiral shift reads as
\begin{eqnarray}
\label{results-sol-Voigt-omega-q-inft-nonrecipr}
\left|\omega_{\pm}(q\to\infty)-\omega_{\pm}(q\to-\infty)\right|&\approx& \frac{d \omega_b \varepsilon_{\infty}\left[\varepsilon_{\infty} \omega_b^2 +4\Omega_e^2(1+\varepsilon_{\infty}) +\omega_b\sqrt{\varepsilon_{\infty}} \sqrt{\varepsilon_{\infty} \omega_b^2 +4\Omega_e^2(1+\varepsilon_{\infty})}\right]}{bl^2 \left\{1+\varepsilon_{\infty} \left[\varepsilon_{\infty}+\omega_b^2 +4\Omega_e^2 (1+\varepsilon_{\infty})\right]\right\}} \nonumber\\
&\approx& \frac{d \omega_b \varepsilon_{\infty}}{bl^2 (1+\varepsilon_{\infty})} = \frac{2e^3 \tilde{A}_{5,y}}{\pi \hbar^2 c(1+\varepsilon_{\infty})}.
\end{eqnarray}
It grows with the magnitude of strain.

\begin{figure}[t]
	\centering
	\subfigure[]{\includegraphics[height=0.275\textwidth]{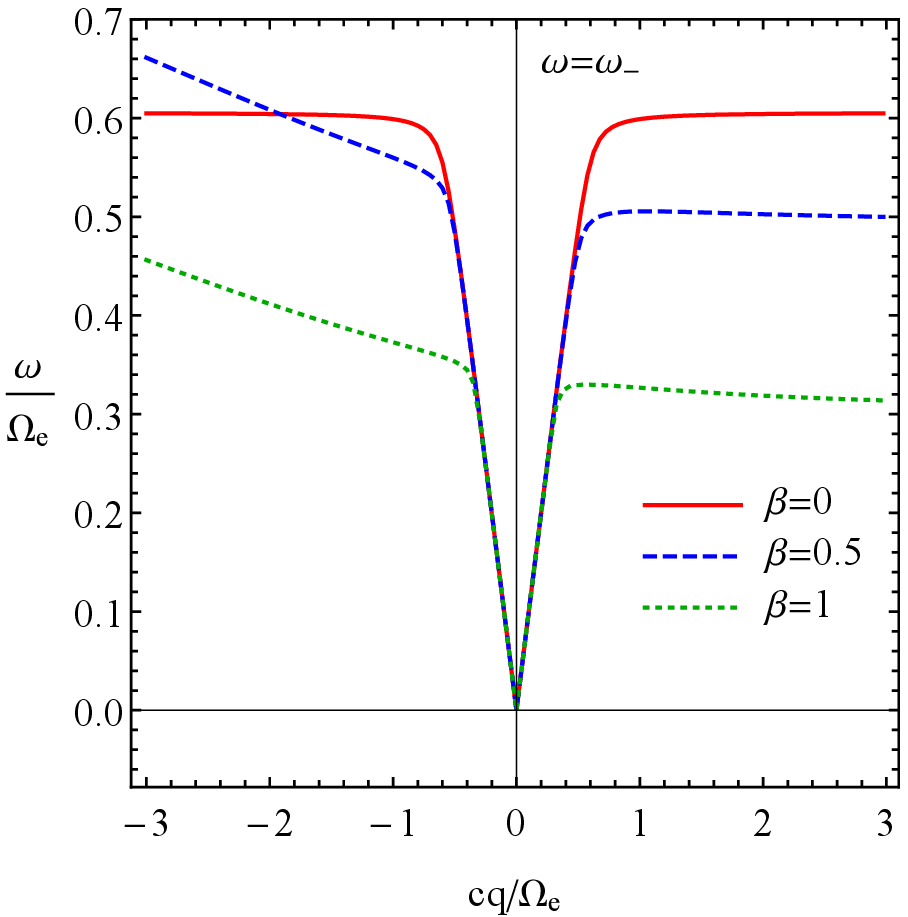}}
	\hspace{0.05\textwidth}
	\subfigure[]{\includegraphics[height=0.275\textwidth]{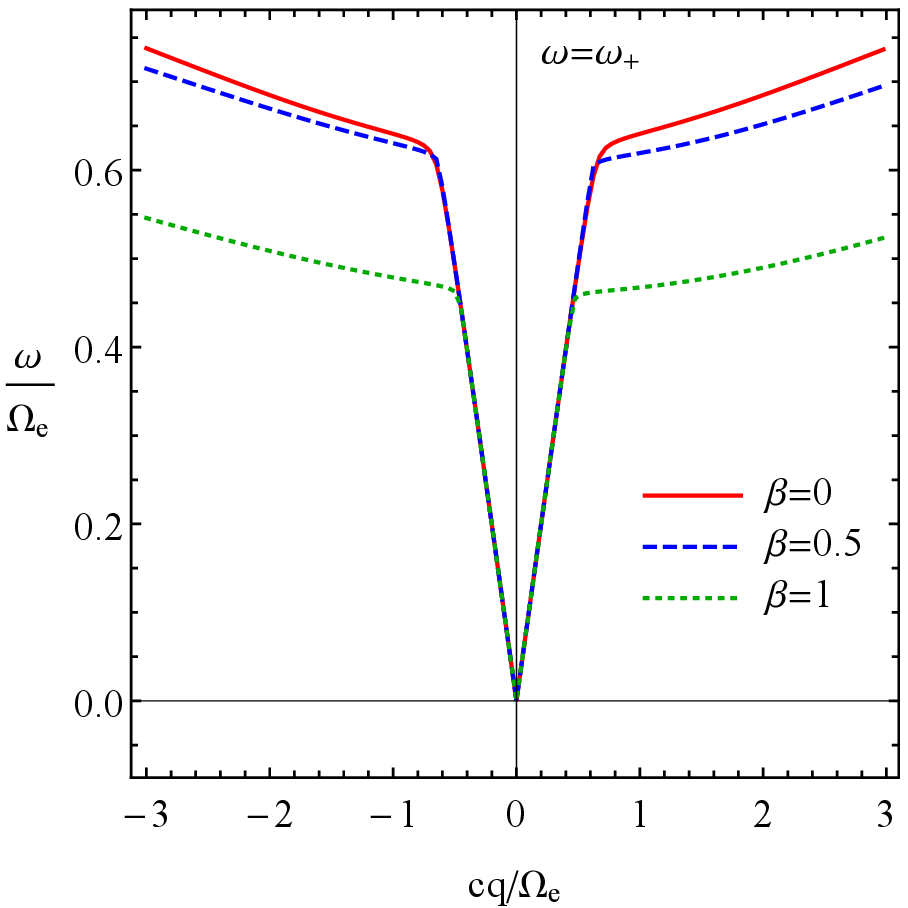}}
	\hspace{0.05\textwidth}
	\subfigure[]{\includegraphics[height=0.275\textwidth]{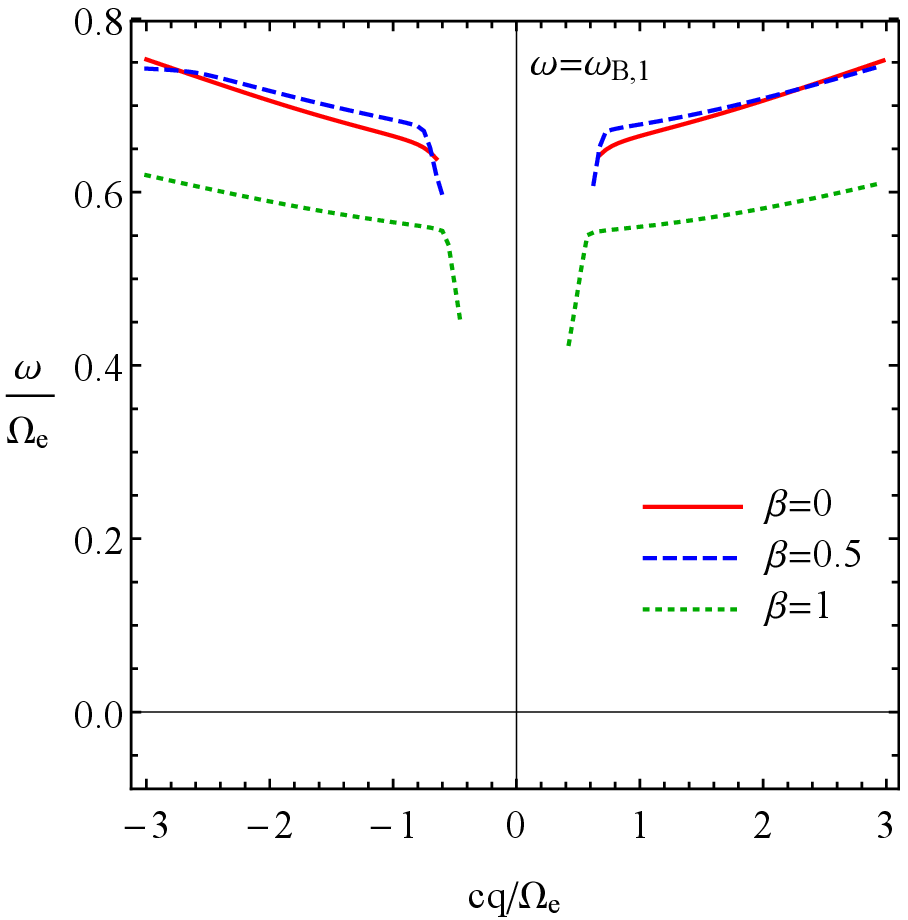}}
	\caption{The dispersion relation of the surface plasmon polaritons in a slab of Weyl semimetal for the Voigt configuration at $\beta=0$ (red solid lines), $\beta=0.5$ (blue dashed lines), and $\beta=1$ (green dotted lines). Panels (a), (b), and (c) correspond to two SPP branches $\omega_{-}$ and $\omega_{+}$, as well as the lowest bulk mode $\omega_{\rm B,1}$, respectively. We set $d=2c/\Omega_e$, and $\omega_b=\Omega_e$.
    }
	\label{fig:results-Voigt-omega-few-beta}
\end{figure}

Numerical results for the SPP dispersion at a few values of the strain strength $\beta$ are shown in Fig.~\ref{fig:results-Voigt-omega-few-beta}. The nonreciprocity of the surface collective modes is clearly evident at large values of strain quantified by $\beta$ and agrees well with the results in Eqs.~(\ref{results-sol-Voigt-omega-q-inft-plus}) and (\ref{results-sol-Voigt-omega-q-inft-minus}).
The nonreciprocity of the SPPs originates from the broken parity-inversion symmetry $z \to -z$ and the Weyl node separation. In the case under consideration, a nonuniform strain breaks this symmetry leading to the dependence of the frequencies on the sign of the SPP wave vector $q$. It is worth noting that the parity-inversion symmetry could be broken also when the slab of an unstrained Weyl semimetal is surrounded by dielectrics with different dielectric constant (see, e.g., Ref.~\cite{Kotov-Lozovik:2018}). Therefore, while the strain is not equivalent to the nonuniform dielectric constant of the sample, its effect on the SPPs appears to be qualitatively similar. The same analogy might be used to explain the decrease of the frequencies at intermediate $q$.

\begin{figure}[t]
	\centering
	\subfigure[]{\includegraphics[height=0.35\textwidth]{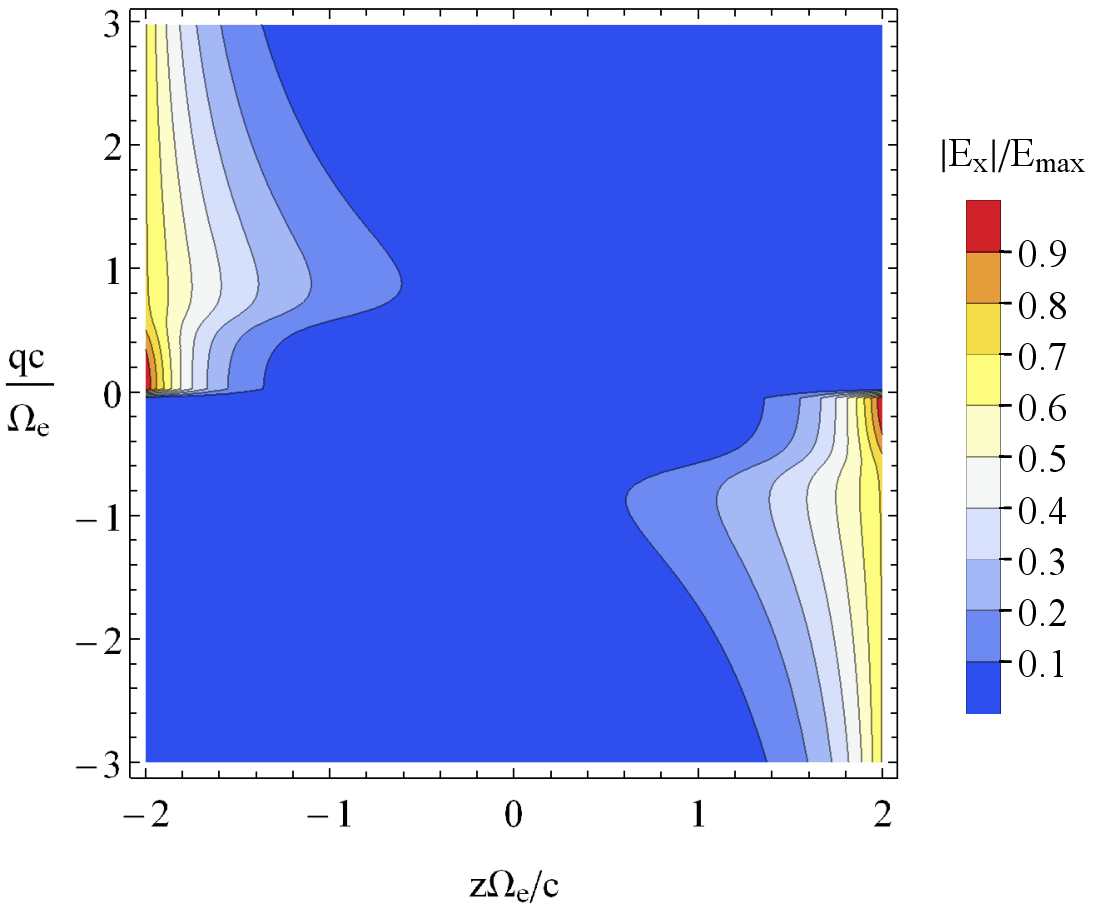}}
	\hspace{0.02\textwidth}
	\subfigure[]{\includegraphics[height=0.35\textwidth]{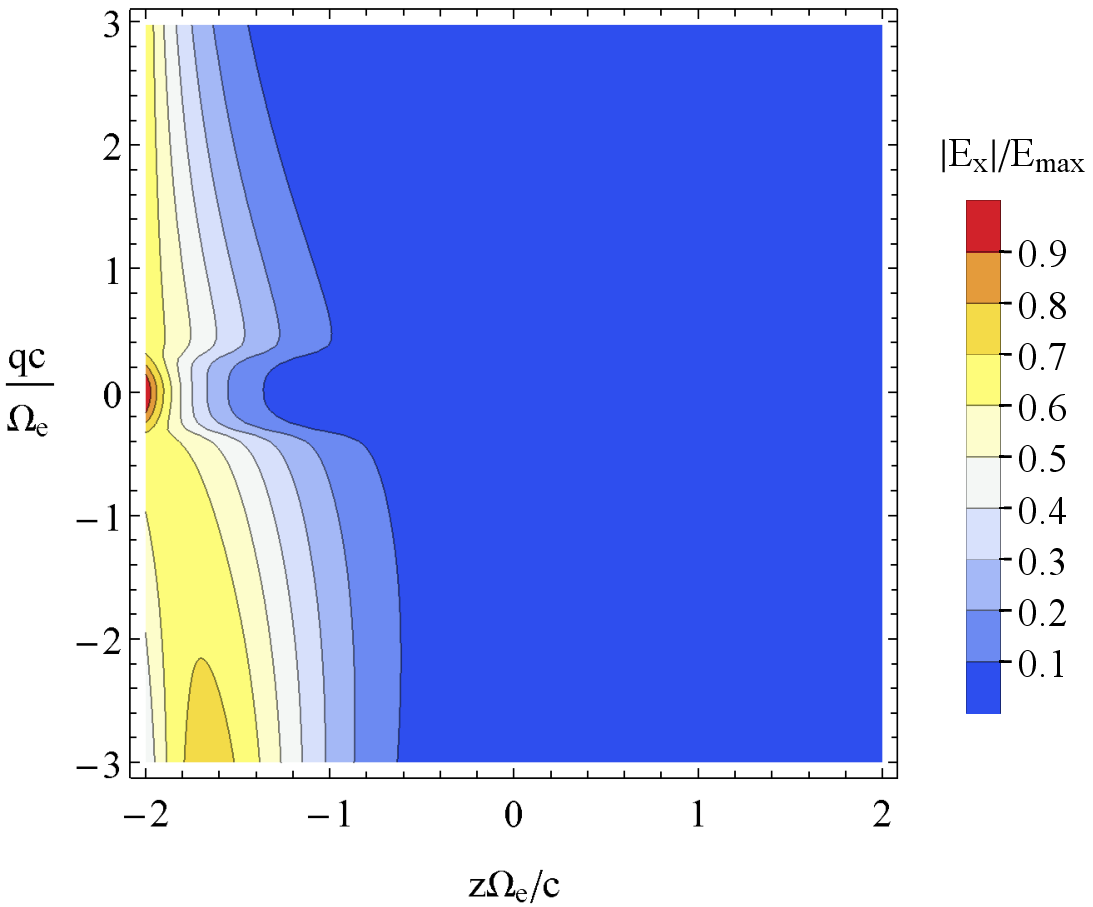}}
	\hspace{0.02\textwidth}
	\subfigure[]{\includegraphics[height=0.35\textwidth]{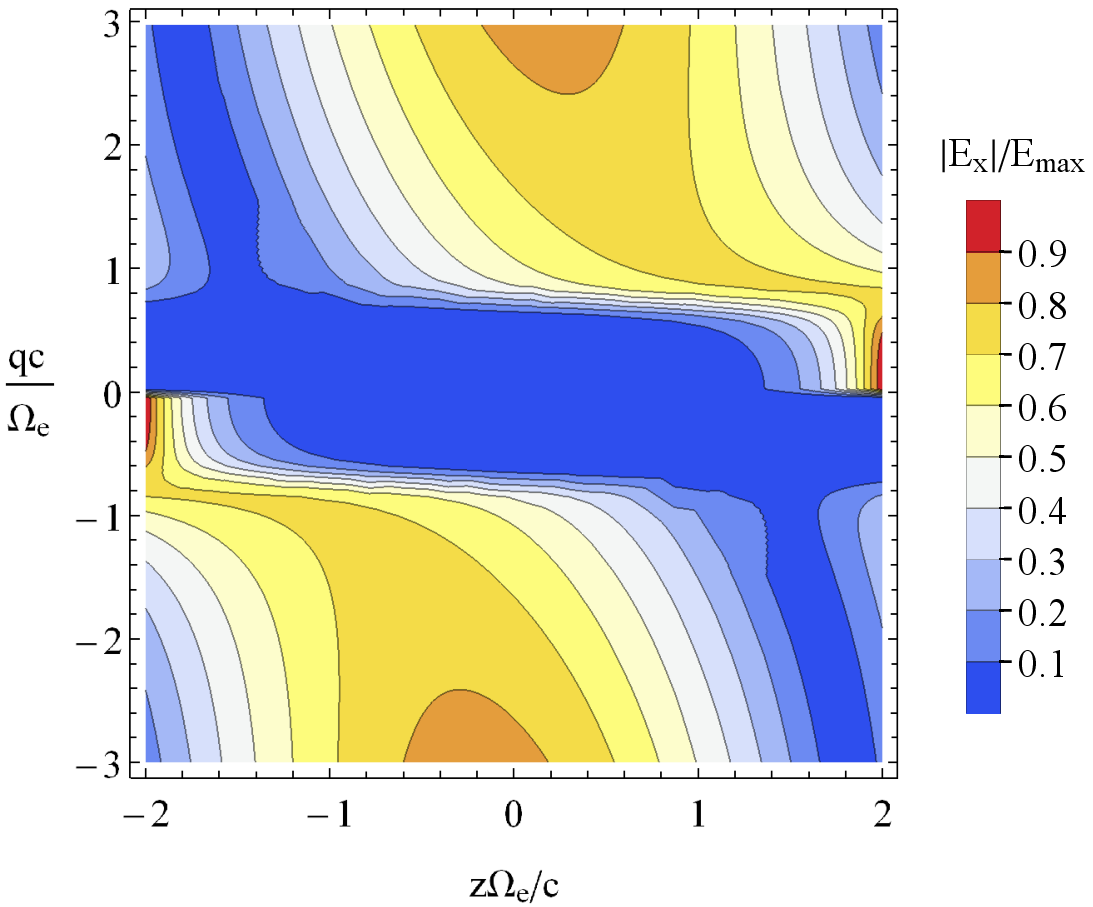}}
	\hspace{0.02\textwidth}
	\subfigure[]{\includegraphics[height=0.35\textwidth]{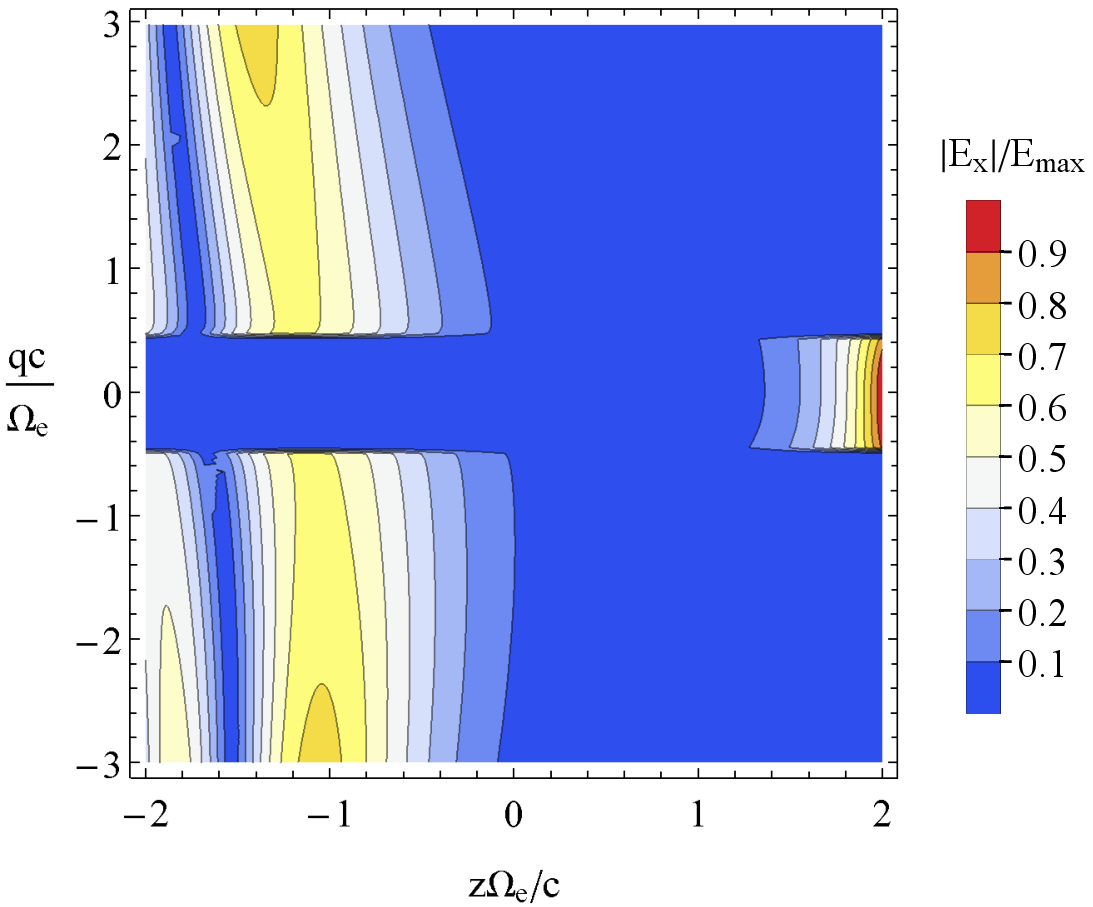}}
	\caption{Profiles of the $x$ component of the electric field $E_x$ in the Voigt configuration. Top and bottom panels correspond to two SPP branches $\omega_{-}$ and $\omega_{+}$, respectively.  Strain strength is $\beta=0$ in panels (a) and (c) as well as $\beta=1$ in panels (b) and (d). We set $d = 2c/\Omega_{e}$ and $\omega_b = \Omega_e$.}
	\label{fig:results-Voigt-fields}
\end{figure}

The spatial distribution of the electric field inside the slab is shown in Fig.~\ref{fig:results-Voigt-fields}. The surface localization of the lowest mode is clearly evident from the figure. On the other hand, the field of the second mode $\omega_{+}$ could become noticeable inside the slab. We checked that the localization become much more pronounced in larger samples. It is worth noting also that the change of the spatial dependence of the field distributions from the exponentially localized to oscillating one can be easily inferred by using Eqs.~(\ref{model-electric-field-ansatz}) and (\ref{results-sol-Voigt-kappaV}) in the case $A_5=0$. Indeed, the parameter $\kappa_V$ is real and positive in the case of surface modes. On the other hand, the mixing with bulk modes leads to an imaginary part of $\kappa_V$. In the strained case, however, one can rely on the spatial profiles of the fields. As one can see from Fig.~\ref{fig:results-Voigt-fields}, there are no purely surface collective modes in the slab because there is always a finite overlap between the surfaces. The localization length, however, depends on the wave vector. Indeed, it is smallest at small wave vectors and tends to increase with $q$.

\subsection{Faraday configuration}
\label{sec:results-sol-Faraday}

Finally, we consider the Faraday configuration. It is schematically shown in Fig.~\ref{fig:setup}(c), where we set $\mathbf{q}\parallel\mathbf{b}\parallel\mathbf{A}_5\parallel \hat{\mathbf{x}}$. The corresponding strain corresponds to bending about the $y$ axis producing $\mathbf{A}_5 \propto u_0b_xz \hat{\mathbf{x}}/d$.

By using Eq.~(\ref{results-wave-eq-expl}) and Gauss's law $\bm{\nabla}\cdot \mathbf{D}=0$, we derive the following equation for $E_y$:
\begin{eqnarray}
\label{results-sol-Faraday-eq}
&&\frac{1}{\varepsilon_2\left[1-z/(bl^2)\right]} E_y^{(4)} + \frac{2}{\varepsilon_2 bl^2\left[1-z/(bl^2)\right]^2} E_y^{(3)} %\nonumber\\&&
+\frac{1}{\varepsilon_1 \varepsilon_2 \left[1-z/(bl^2)\right]^3} \Bigg\{\varepsilon_1 \left[\frac{2}{(bl^2)^2} -(q^2 +\kappa^2) \left(1-\frac{z}{bl^2}\right)^2\right] \nonumber\\
&&+\frac{\omega^2}{c^2} \left(1-\frac{z}{bl^2}\right)^2 \left[(\varepsilon_1-\varepsilon_2) +\frac{z}{bl^2} \varepsilon_2\right] \left(\varepsilon_1 +\varepsilon_2 -\varepsilon_2 \frac{z}{bl^2}\right)\Bigg\}E_y^{\prime \prime} %\nonumber\\&&
+\frac{\omega^2}{c^2 bbl^2}\left[\frac{2\varepsilon_2}{\varepsilon_1} -\frac{c^2\kappa^2}{\varepsilon_2\omega^2 \left[1-z/(bl^2)\right]^2}\right] E_y^{\prime} \nonumber\\
&&-\frac{1}{\varepsilon_2 \left[1-z/(bl^2)\right]^3} \left\{\kappa^2\left[\frac{2}{(bl^2)^2} -q^2\left(1-\frac{z}{bl^2}\right)^2\right] +\frac{\omega^2}{c^2} \kappa^2 \varepsilon_1 \left(1-\frac{z}{bl^2}\right)^2 +\varepsilon_2^2 \frac{\omega^4}{c^4} \left(1-\frac{z}{bl^2}\right)^4
\right\}E_y=0.
\end{eqnarray}
The $z$ and $x$ components of the electric field are determined by
\begin{eqnarray}
\label{results-sol-Faraday-Ez}
E_z &=& c^2\frac{\kappa^2 E_y - E_y^{\prime \prime}}{i\varepsilon_2\omega^2\left[1-z/(bl^2)\right]},\\
\label{results-sol-Faraday-Ex}
E_x &=& -\frac{\varepsilon_1 E_z^{\prime}+ i\varepsilon_2 E_y/(bl^2) -i\varepsilon_2 \left[1-z/(bl^2)\right] E_y^{\prime}}{iq \varepsilon_1},
\end{eqnarray}
respectively.

The decay constant $\kappa_{F}$ can be obtained analytically at $l \rightarrow \infty$ and $d\to\infty$. It reads as
\begin{equation}
\label{results-sol-Faraday-kappa}
\kappa_{F}^2 = q^2 + \frac{\omega^2}{c^2}\left(\frac{\varepsilon_2^2}{2\varepsilon_1} - \varepsilon_1\right) \pm \frac{\varepsilon_2 \omega^2}{2c^2 |\varepsilon_1|} \sqrt{\varepsilon_2^2 +\frac{4c^2 q^2 \varepsilon_1}{\omega^2}}.
\end{equation}
This result agrees with that in Ref.~\cite{Hofmann-DasSarma:2016}.

\begin{figure}[t]
	\centering
	\subfigure[]{\includegraphics[height=0.275\textwidth]{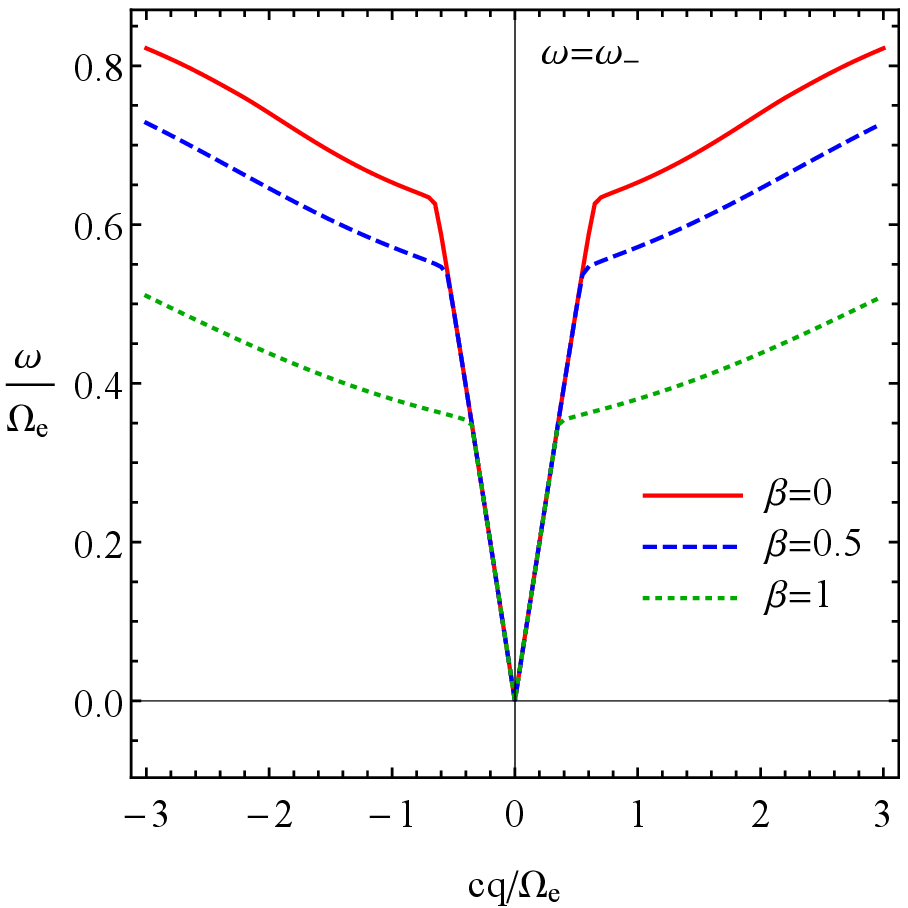}}
	\hspace{0.02\textwidth}
	\subfigure[]{\includegraphics[height=0.275\textwidth]{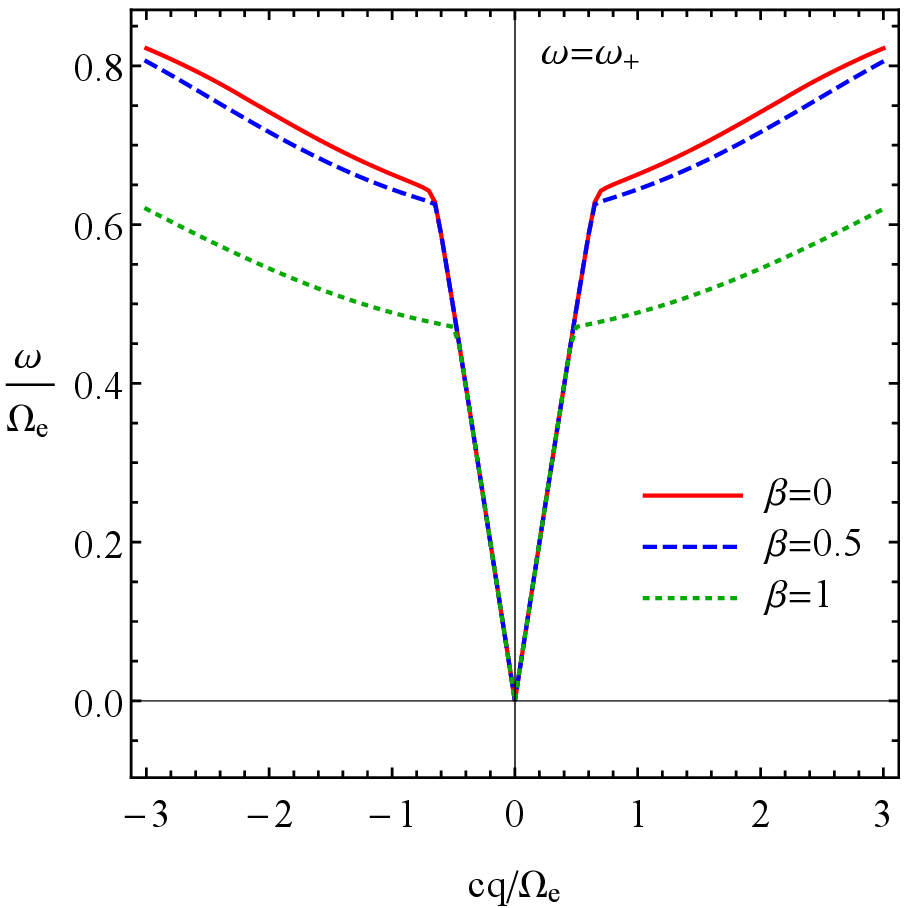}}
	\hspace{0.02\textwidth}
	\subfigure[]{\includegraphics[height=0.275\textwidth]{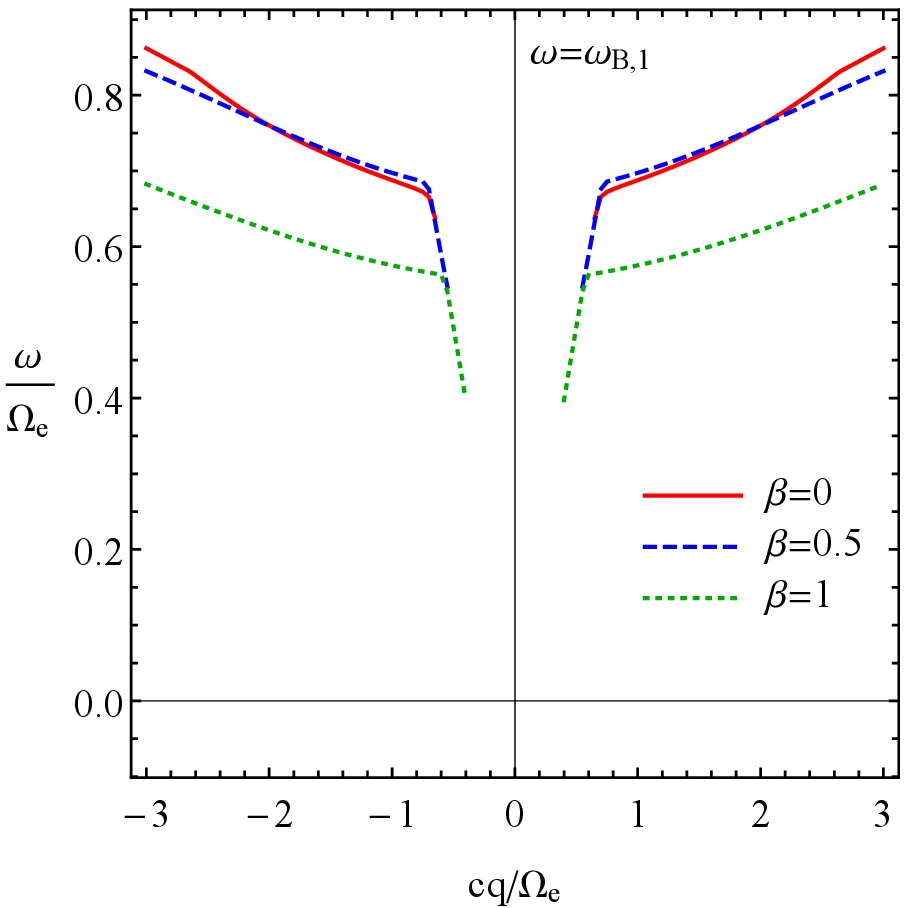}}
	\caption{Dispersion relation of the collective modes in a slab of Weyl semimetal for the Faraday configuration at $\beta=0$ (red solid lines), $\beta=0.5$ (blue dashed lines), and $\beta=1$ (green dotted lines). Panels (a), (b), and (c) correspond to two SPP branches $\omega_{-}$ and $\omega_{+}$, as well as the lowest bulk mode $\omega_{\rm B,1}$, respectively. We set $d=2c/\Omega_e$, and $\omega_b=\Omega_e$.
    }
	\label{fig:results-Faraday-omega-few-beta}
\end{figure}

Let us analyze the analytical solutions at small and large wave vectors. The dispersion relation is the same as in the other two configurations (see Secs.~\ref{sec:results-sol-perpendicular} and \ref{sec:results-sol-Voigt}), i.e., $\omega=cq$ at small wave vectors. In the case $q \to \pm\infty$, Eq.~(\ref{results-sol-Faraday-eq}) simplifies
\begin{equation}
\label{results-sol-Faraday-eq-large-q}
E_y^{(4)} + \frac{2 E_y^{(3)}}{b l^2 - z} -2 q^2 E_y^{\prime \prime} - \frac{2 q^2 E_y^{\prime}}{b l^2 - z} + q^4 E_y = 0.
\end{equation}
Its general solution is
\begin{equation}
E_y = C_1 e^{qz} +C_2 e^{-qz} +C_3 e^{qz}z \left[3+qbl^2\left(2-\frac{z}{bl^2}\right)\right] +C_4e^{-qz} z \left[3-qbl^2\left(2-\frac{z}{bl^2}\right)\right].
\end{equation}
By using this solution and employing the continuity relations for $\partial_z E_y$ and $\partial_z E_x - i q E_z$ at the surface, we found that $\omega(q\to\pm \infty)$ are given by the same expression as in Eq.~(\ref{results-sol-Perp-omega-q-inf}). Therefore, the corresponding modes are reciprocal even in the presence of deformations and the chiral shift. We present the dispersion relations of the SPPs in Fig.~\ref{fig:results-Faraday-omega-few-beta} at a few values of strain strength quantified by $\beta$. The effects of strains are similar to those in the perpendicular configuration (see Sec.~\ref{sec:results-sol-perpendicular}).

\begin{figure}[t]
	\centering
	\subfigure[]{\includegraphics[height=0.35\textwidth]{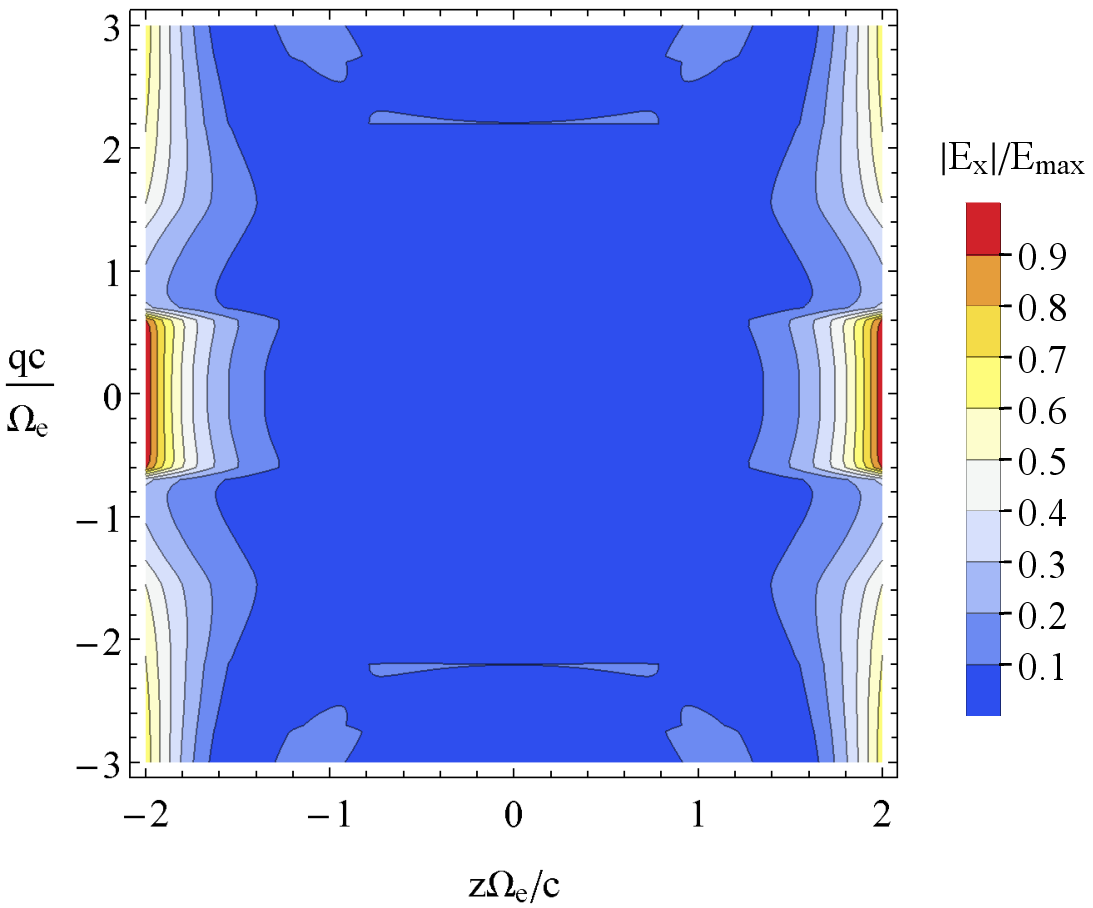}}
	\hspace{0.02\textwidth}
	\subfigure[]{\includegraphics[height=0.35\textwidth]{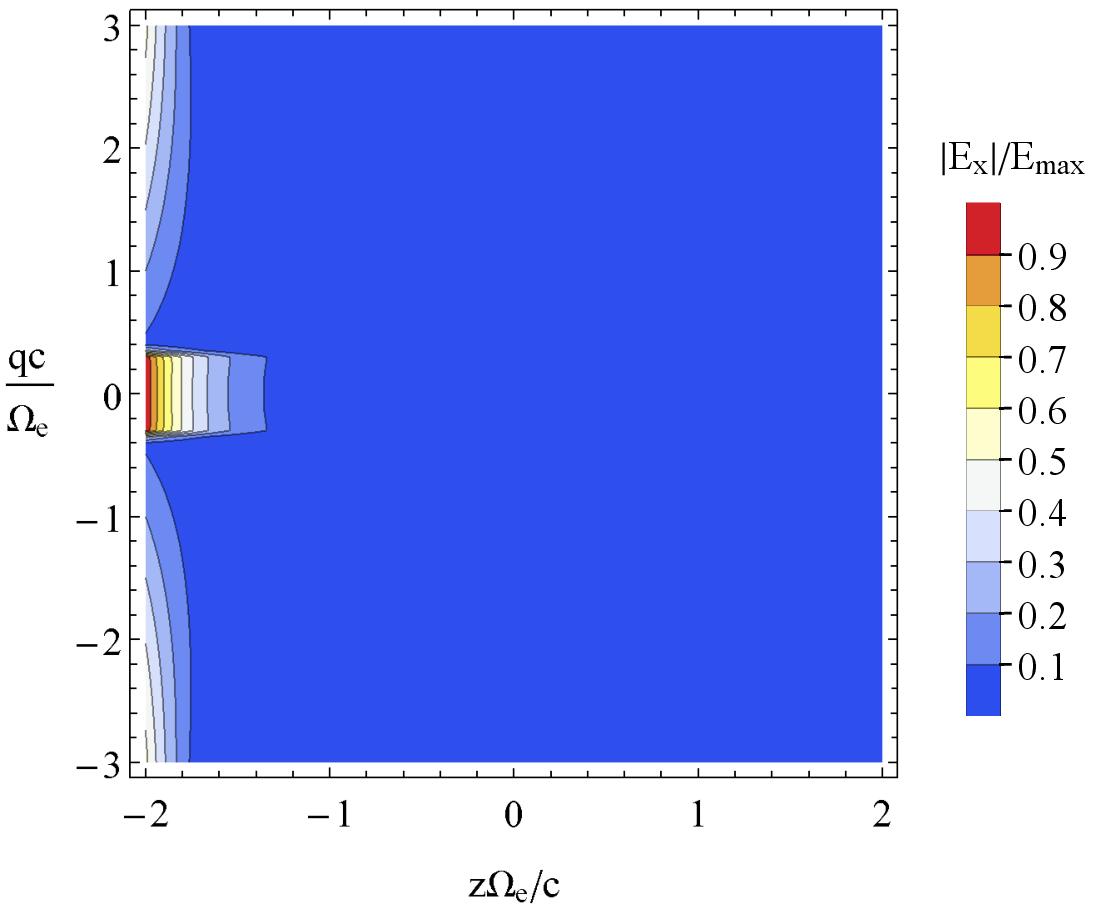}}
	\hspace{0.02\textwidth}
	\subfigure[]{\includegraphics[height=0.35\textwidth]{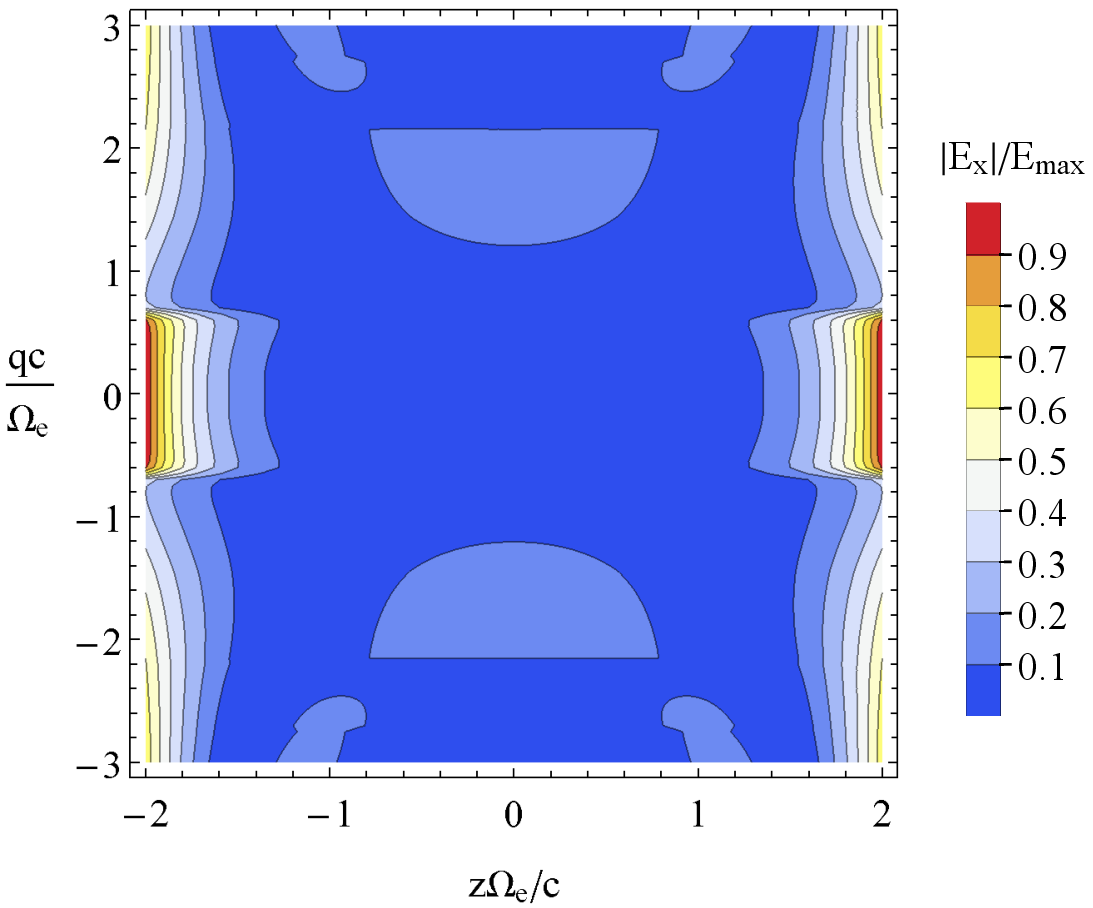}}
	\hspace{0.02\textwidth}
	\subfigure[]{\includegraphics[height=0.35\textwidth]{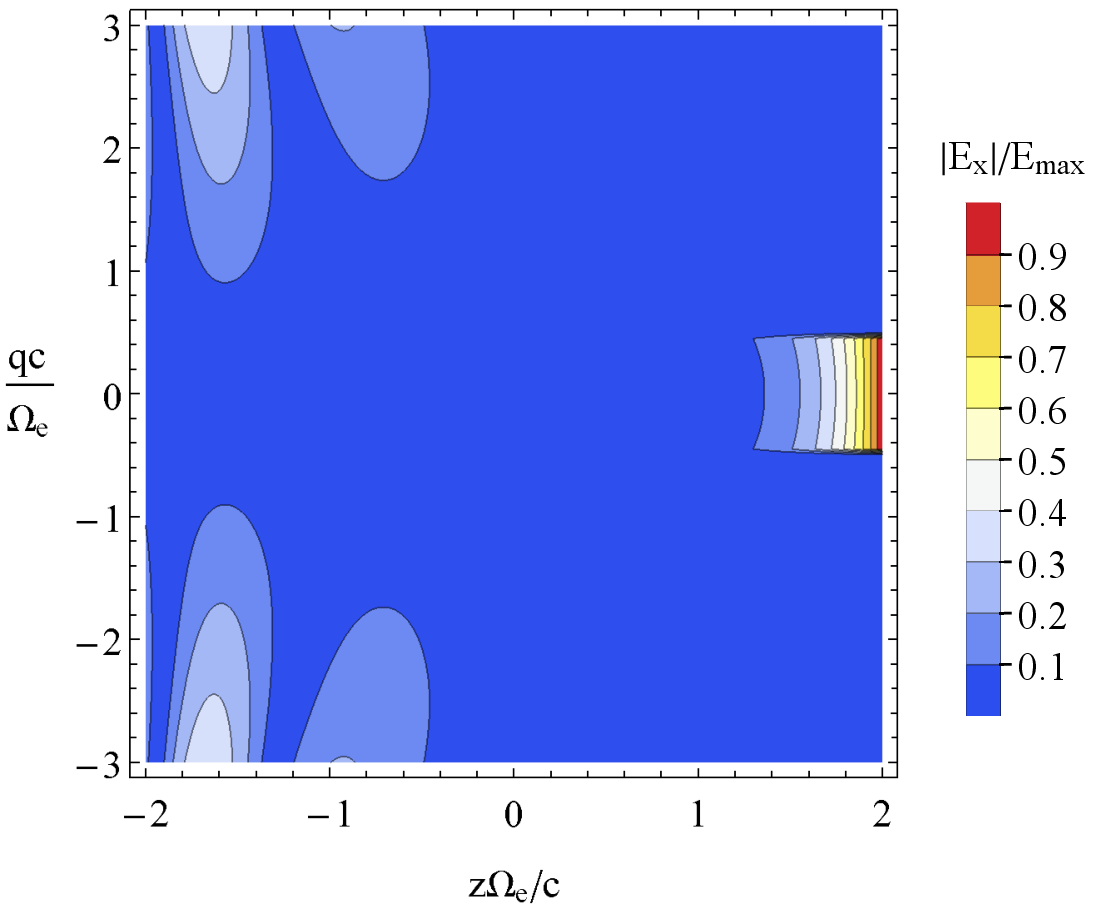}}
	\caption{Profiles of the $x$ component of the electric field $E_x$ in the Faraday configuration. Top and bottom panels correspond to two SPP branches $\omega_{-}$ and $\omega_{+}$, respectively. Strain strength is $\beta=0$ in panels (a) and (c) as well as $\beta=1$ in panels (b) and (d). We set $d = 2c/\Omega_{e}$ and $\omega_b = \Omega_e$.}
	\label{fig:results-Faraday-fields}
\end{figure}

Finally, let us discuss the profiles of electric field. We present the corresponding results in Fig.~\ref{fig:results-Faraday-fields}. As one can see, the lowest mode is well localized for small wave vectors. Strain, however, changes the surface where the mode is localized. Therefore, the lowest mode could be identified with a surface mode or a short-range surface plasmon~\cite{Tamaya-Kawabata:2019}. A similar effect of strain is also present for the second mode $\omega_{+}$. The field magnitude in the bulk is more pronounced in this case, however.

\section{Summary}
\label{sec:Summary}

In this study, we investigated the effects of strains on the surface plasmon polaritons in a Weyl semimetal slab. By using a low-energy model of a time-reversal symmetry broken Weyl semimetal, we found that strain provides an effective means to control the nonreciprocity and localization of the SPPs. As in the previous studies, the collective modes strongly depend on the relative orientation of the chiral shift $\mathbf{b}$, the wave vector $\mathbf{q}$ of collective modes, and the surface normal $\hat{\mathbf{n}}$ for which the three main configurations can be identified. They are the perpendicular ($\mathbf{b}\parallel\hat{\mathbf{n}}$), Voigt ($\mathbf{b}\perp\hat{\mathbf{n}}$ and $\mathbf{b}\perp \mathbf{q}$), and Faraday ($\mathbf{b}\parallel\mathbf{q}$) configurations.

By applying bending and inhomogeneous stretching, a coordinate-dependent axial gauge field that does not break the translation invariance along the surface of the slab can be generated. For the perpendicular and Faraday configurations, this strain-induced field reduces the frequencies of the collective modes for intermediate values of the wave vector $q$ (there is no dependence on strain at $q\to\pm\infty$) and enhances their localization at the surfaces. Moreover, strain can even change the localization of the SPPs introducing an asymmetry in their field profiles. The results for the Voigt configuration demonstrate that the strain-induced axial gauge field generated by bending not only reduces the frequency of the modes but makes the SPPs nonreciprocal even in thin films. The nonreciprocity of the SPPs originates from the separation between the Weyl nodes in momentum space and broken parity-inversion symmetry $z\to -z$ due to a nonuniform strain. This finding is quite interesting since the nonreciprocity is usually absent in slabs of finite thickness due to the hybridization of the collective modes at different surfaces.

The proposed effect could have a direct practical application. Indeed, strain-induced axial gauge fields provide an efficient way to create tunable unidirectional optical devices. Among them, we mention nonreciprocal circulators, nonreciprocal Mach-Zehnder interferometers, and one-way waveguides. Unlike previous proposals, where the thickness of a Weyl semimetal, dielectric constants of surrounding media, and the direction of the chiral shift were used, the nonreciprocity in the proposed setup can be manipulated {\it in situ}.
Numerical estimates suggest that the strain-induced effects could be potentially measured for sufficiently high strain magnitude and thin films. Experimentally, strain-induced modifications of SPPs could be realized, for example, in the recently discovered Weyl semimetal EuCd$_2$As$_2$, where only two Weyl nodes separated in momentum space exist in the vicinity of the Fermi level~\cite{Soh-Boothroyd:2019,Ma-Shi:2019}.

Finally, let us comment on the nonuniform profile of the chiral shift that is realized at the surface of Weyl semimetals (see Appendix~\ref{sec:chiral-shift}). Contrary to external strains, where the chiral shift profile is asymmetric inside the slab, a symmetric profile reduces the localization of the surface collective modes. While the Weyl node separation is always nonuniform in finite samples of Weyl semimetals, the corresponding modification of the anomalous Hall conductivity is estimated to be weak.

\begin{acknowledgments}
The work of E.V.G. was supported partially by the National Academy of Sciences of Ukraine grants No.~0116U003191 and No.~0120U100858. P.O.S. was supported partially by the VILLUM FONDEN via the Centre of Excellence for Dirac Materials (Grant No.~11744), the European Research Council under the European Unions Seventh Framework Program Synergy HERO, and the Knut and Alice Wallenberg Foundation KAW 2018.0104.
\end{acknowledgments}

\appendix

\section{Effects of nonuniform chiral shift profile}
\label{sec:chiral-shift}

\begin{figure}[t]
	\centering
\includegraphics[height=0.35\textwidth]{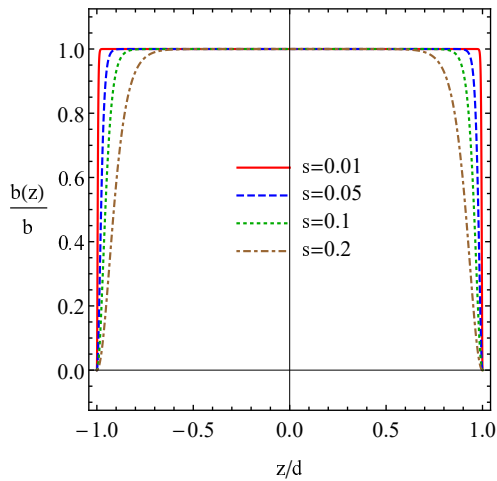}
	\caption{Schematic profile of the chiral shift defined in Eq.~(\ref{chiral-shift-bz-tanh}) for a few values of the profile curvature $s$ at $d=2c/\Omega_e$.
    }
	\label{fig:chiral-shift-profile}
\end{figure}

In addition to external strain, a nonuniform profile of the chiral shift $\mathbf{b}$ is always present at surfaces of Weyl semimetals. Indeed, the chiral shift is intrinsically nonuniform in a finite slab of a Weyl semimetal because the shift vanishes at the surface (see also the discussion in Sec.~\ref{sec:model}). Mathematically, a nonuniform profile of $\mathbf{b}$ can be modeled as
\begin{equation}
\label{chiral-shift-bz-tanh}
\mathbf{b}(z) = \mathbf{b}\left\{\tanh^2{\left[\frac{(z + d)\Omega_e}{s c}\right]} + \tanh^2{\left[\frac{(z - d)\Omega_e}{s c}\right]}- \tanh^2{\left(\frac{2 d \Omega_e}{sc}\right)} \right\}.
\end{equation}
Here, parameter $s$ defines the curvature of the chiral shift profile. The chiral shift is uniform inside the slab, $\mathbf{b}(z)\to \mathbf{b} \theta(|z|-d)$, in the limit $s\to0$ and gradually develops a nonzero curvature for large $s$. The profile given in Eq.~(\ref{chiral-shift-bz-tanh}) is shown schematically in Fig.~\ref{fig:chiral-shift-profile} for several values of $s$.

\begin{figure}[t]
	\centering
	\subfigure[]{\includegraphics[height=0.275\textwidth]{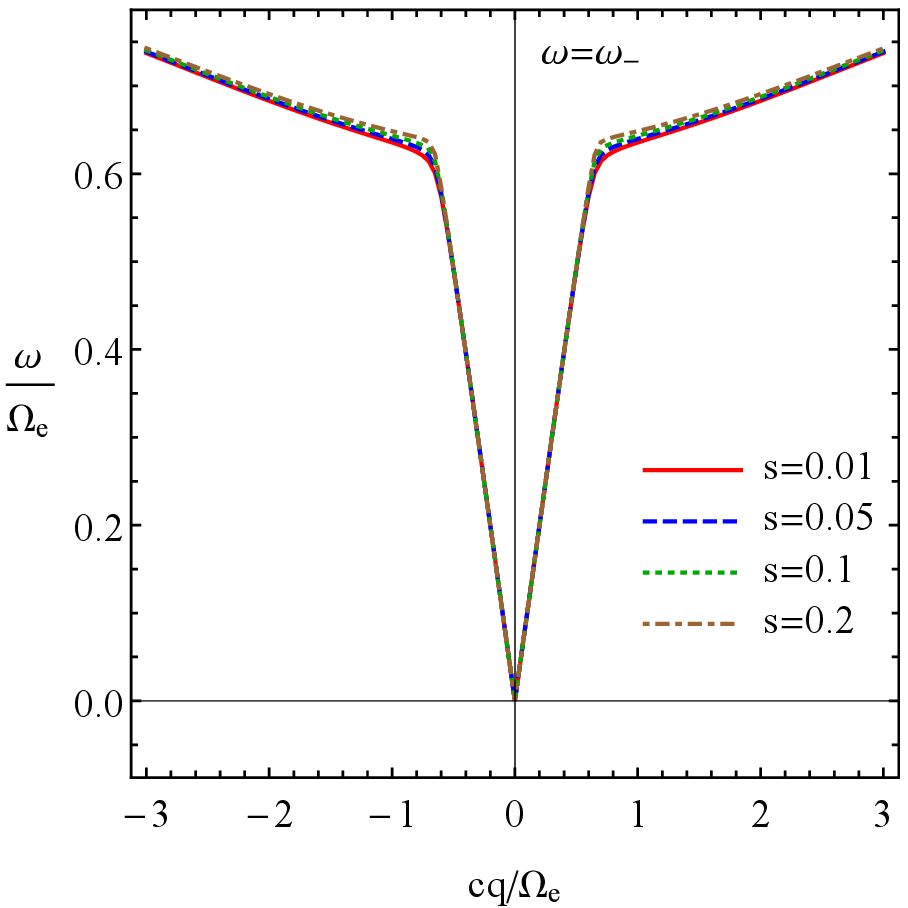}}
	\hspace{0.02\textwidth}
	\subfigure[]{\includegraphics[height=0.275\textwidth]{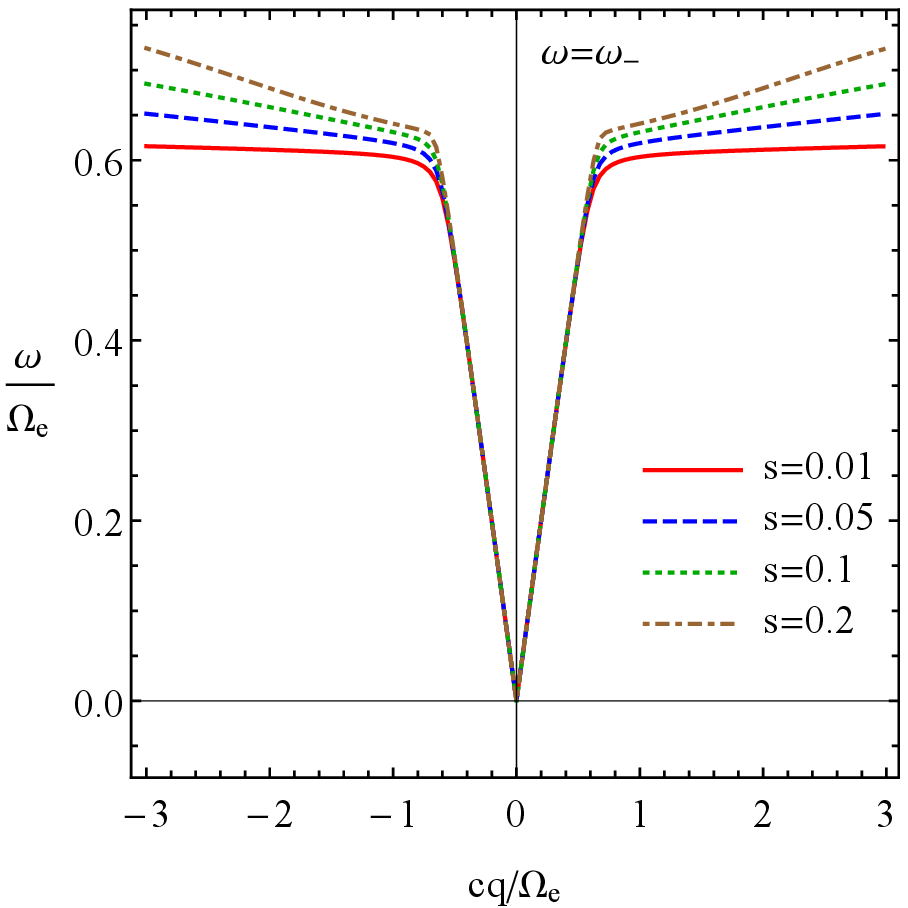}}
	\hspace{0.02\textwidth}
	\subfigure[]{\includegraphics[height=0.275\textwidth]{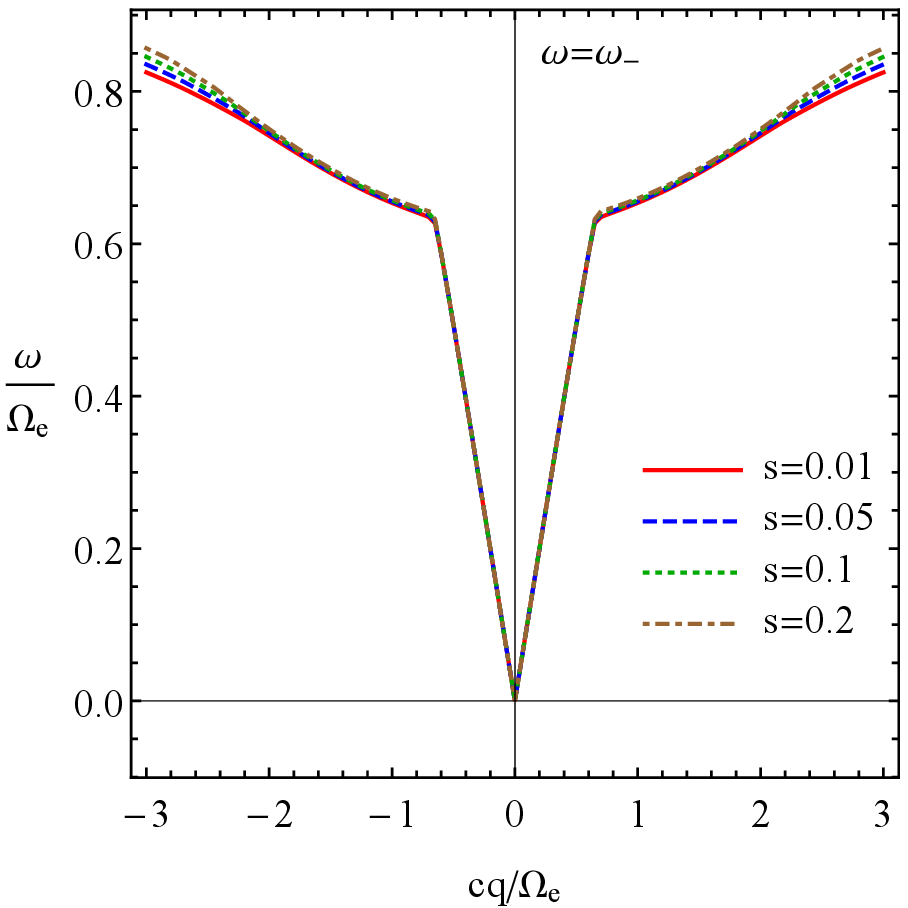}}
	\caption{Dispersion relation of the lowest SPP branch $\omega_{-}$ for the perpendicular (panel (a)), Voigt (panel (b)), and Faraday (panel (c)) configurations at a few values of the chiral shift profile curvature $s$. We set $d=2c/\Omega_e$, and $\omega_b=\Omega_e$.
    }
	\label{fig:chiral-shift-lowest-branch}
\end{figure}

\begin{figure}[t]
	\centering
	\subfigure[]{\includegraphics[height=0.25\textwidth]{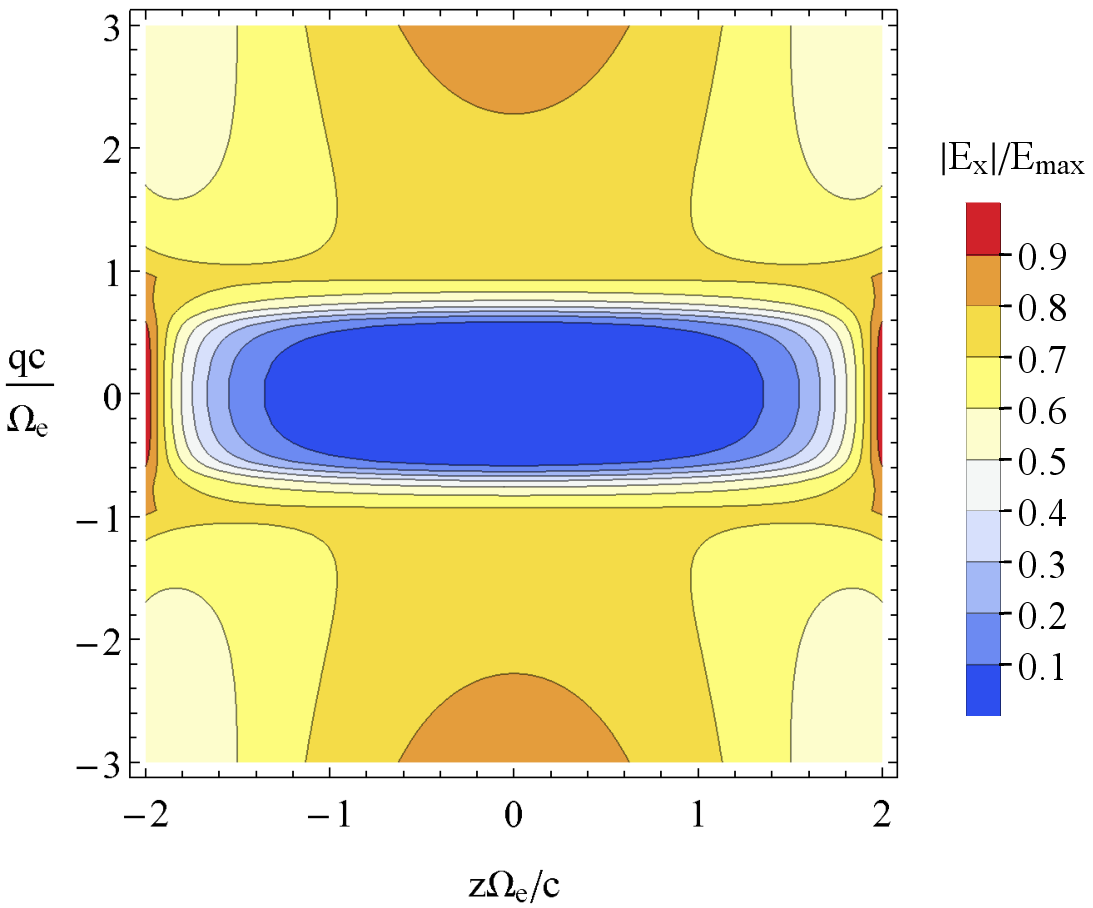}}
	\hspace{0.02\textwidth}
	\subfigure[]{\includegraphics[height=0.25\textwidth]{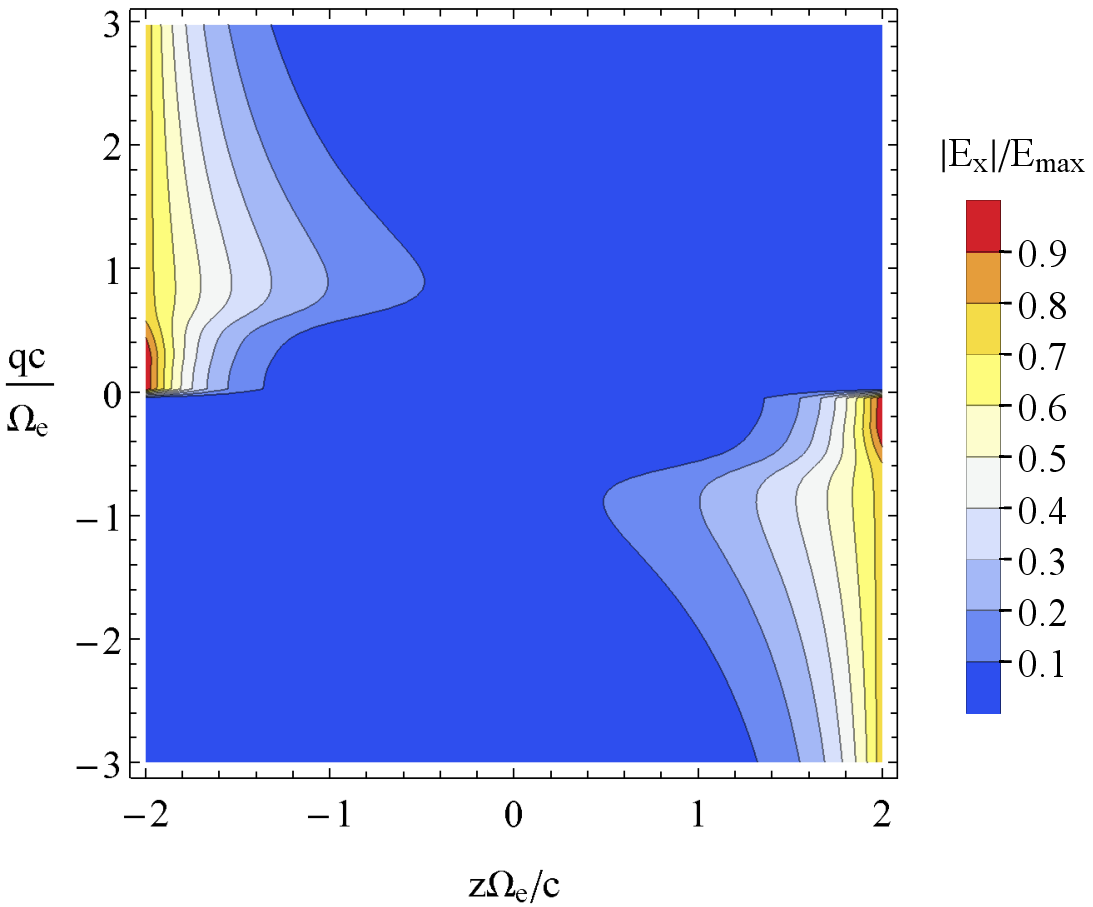}}
	\hspace{0.02\textwidth}
	\subfigure[]{\includegraphics[height=0.25\textwidth]{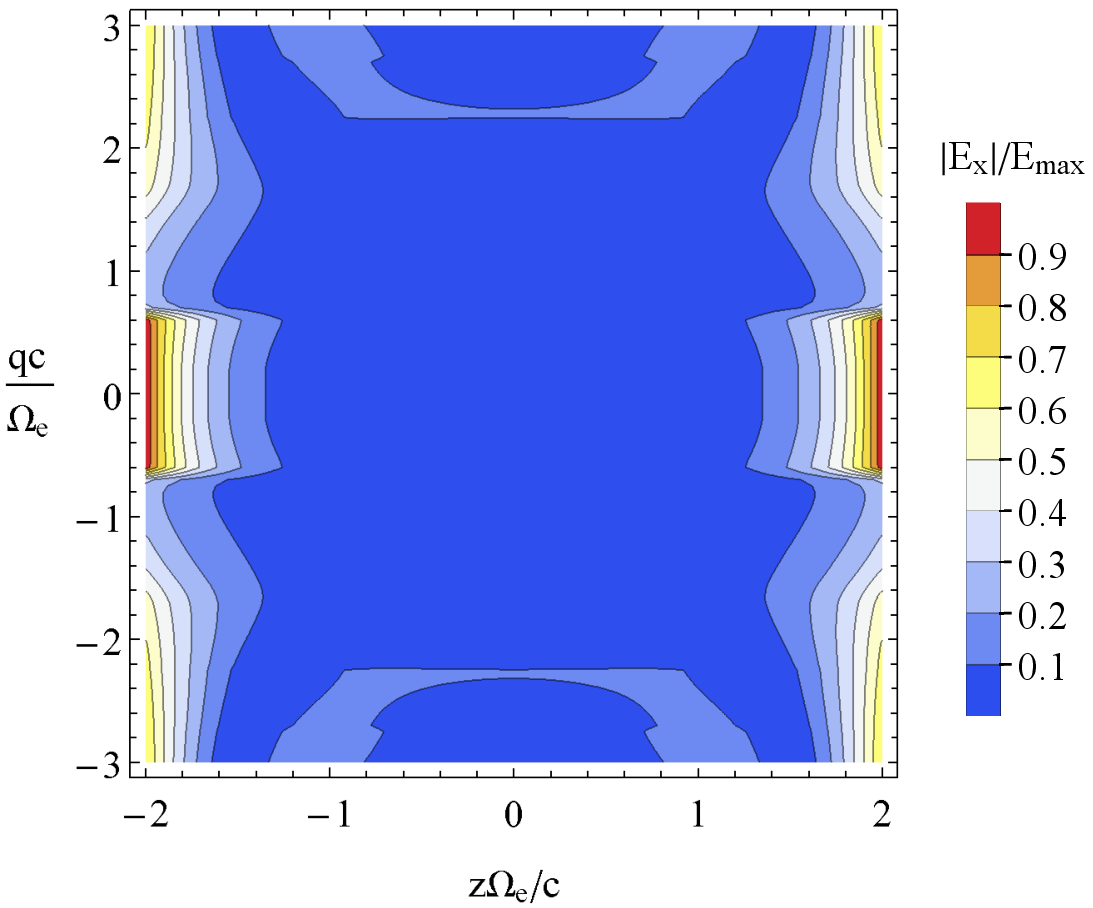}}
	\hspace{0.02\textwidth}
	\subfigure[]{\includegraphics[height=0.25\textwidth]{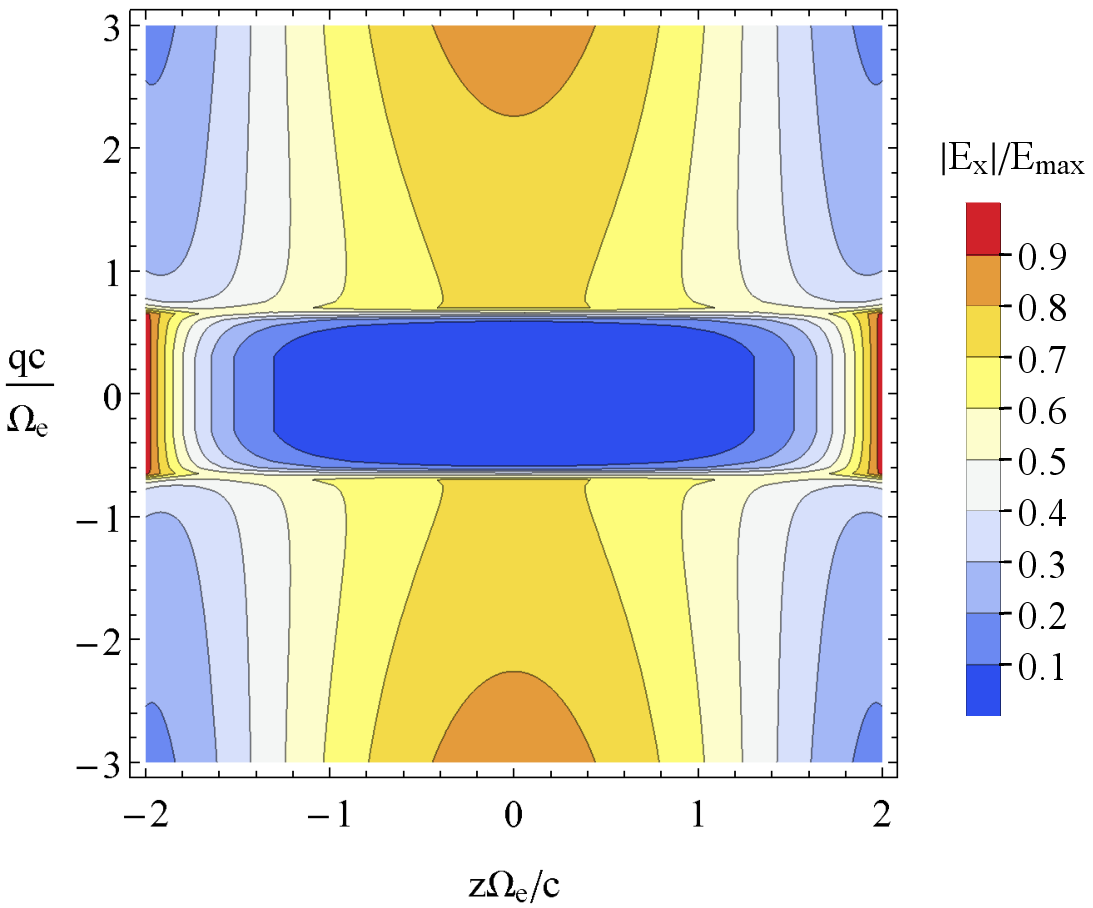}}
	\hspace{0.02\textwidth}
	\subfigure[]{\includegraphics[height=0.25\textwidth]{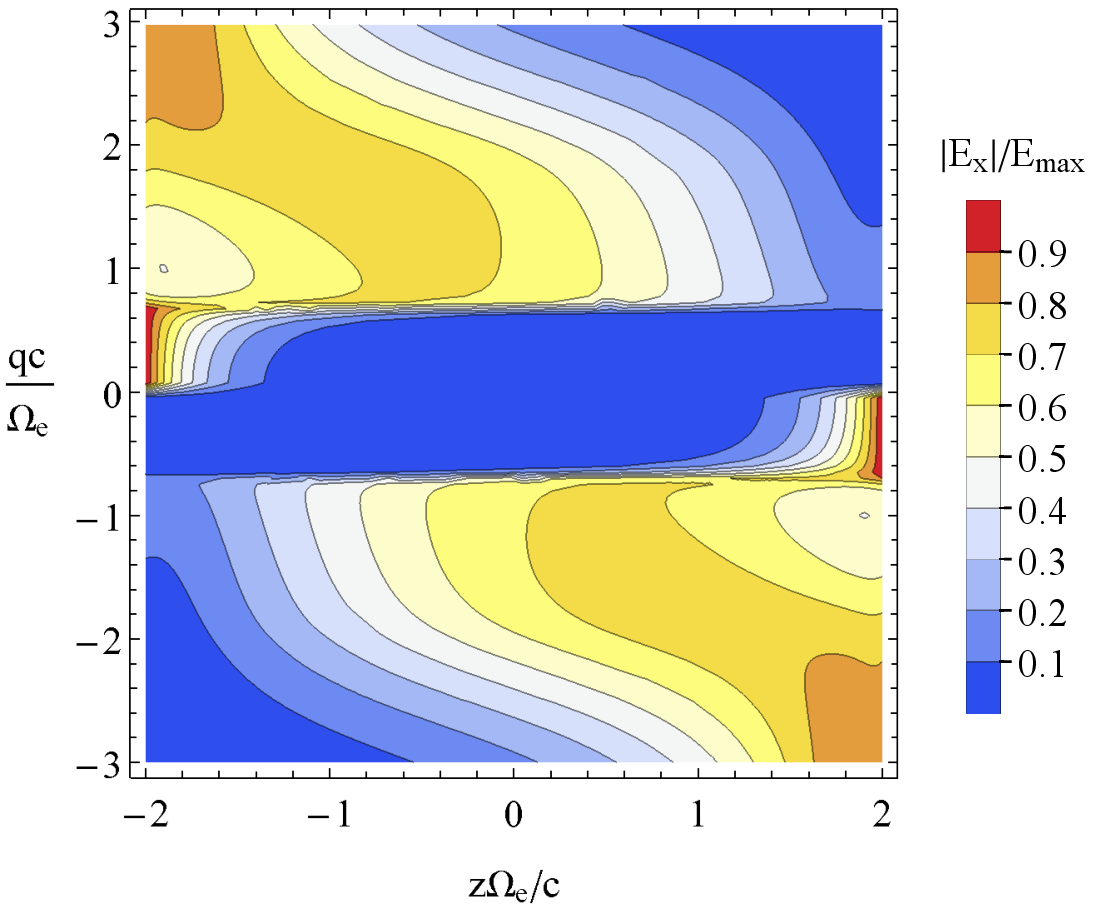}}
	\hspace{0.02\textwidth}
	\subfigure[]{\includegraphics[height=0.25\textwidth]{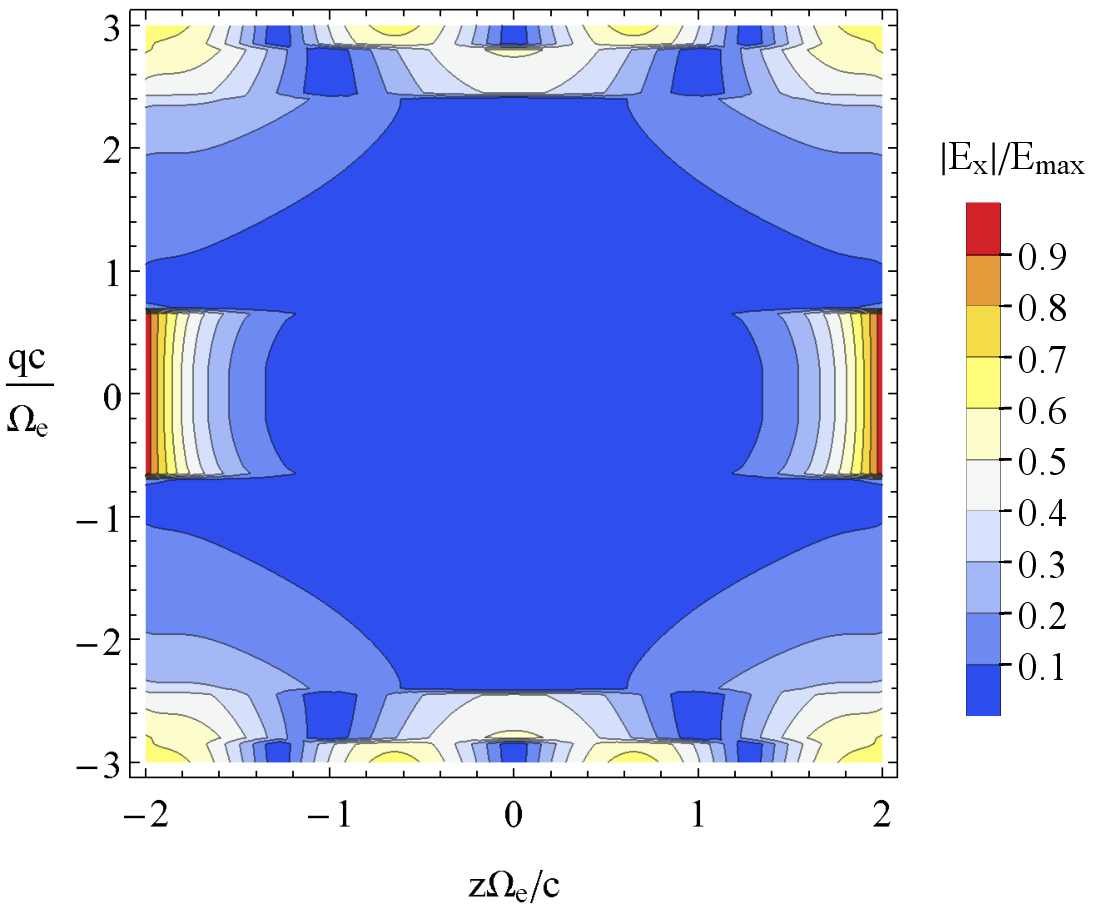}}
	\caption{Profiles of the $x$ component of the electric field $E_x$ for the lowest SPP branch $\omega_{-}$ in the perpendicular (panels (a) and (d)), Voigt (panels (b) and (e)), Faraday (panels (c) and (f)) configurations. Top and bottom panels correspond to steep $s=0.01$ and curved $s=0.2$ profiles of the chiral shift, respectively. We set $d=2c/\Omega_e$, and $\omega_b=\Omega_e$.}
	\label{fig:chiral-shift-field-profiles}
\end{figure}

The calculation of dispersion relations of the collective mode and the corresponding electric field distributions can be performed along the same lines as in Sec.~\ref{sec:results}. Technically, one needs to replace $1-z/(bl^2)$ with $b(z)$. Therefore, we present and discuss only the final results. The spectrum of the lowest SPP branch for the perpendicular, Voigt, and Faraday configurations is shown in Fig.~\ref{fig:chiral-shift-lowest-branch}. In general, a nonzero curvature of the chiral shift profile increases the frequencies of surface plasmon polaritons and bulk modes. It is interesting that the lowest branches of the SPPs are the most susceptible to the nonuniform $\mathbf{b}(z)$. In addition, the results depend on the configuration. For example, the most pronounced effect of the nonuniform chiral shift profile occurs for the Voigt configuration. Furthermore, as one can see by comparing the top and bottom panels of Fig.~\ref{fig:chiral-shift-field-profiles}, a large curvature $s$ of the Weyl node separation profile reduces the localization of the collective modes. This effects is clearly noticeable for the Voigt configuration shown in Figs.~\ref{fig:chiral-shift-field-profiles}(b) and (e). These results suggest that Weyl semimetals might be intrinsically more prone to the delocalization of the surface collective modes. On the other hand, axial gauge field induced by external strains can easily overcome the corrections due to a nonuniform chiral shift profile. It is important also that, unlike surface-induced intrinsic profile, strain-induced axial gauge fields can be easily tuned.

Let us estimate whether the effect is important in real materials. Since the position of the Weyl nodes in momentum space derived from the analysis of the surface Fermi arc states agrees well with the results of the bulk measurements (see, e.g., Refs.~\cite{Yan-Felser:2017-Rev,Hasan-Huang:rev-2017,Armitage-Vishwanath:2017-Rev}), it is unlikely that the characteristic length scale of the chiral shift profile exceeds a few nanometers. On the other hand, our calculations suggest that for the effects of an nonuniform profile to be noticeable, the length scale should be about $0.1d\approx45~\mbox{nm}$. Therefore, we conclude that the intrinsic nonuniform profile of the chiral shift is unlikely to have a profound effect on the surface collective modes.

\end{document}